\documentclass[traditabstract]{aa}

\usepackage{graphicx}
\usepackage{natbib}
\usepackage{lscape}
\usepackage{supertabular}
\usepackage{xcolor}

\newcommand{\vsini}{$v \sin{i}$}

\newcommand{\kir}[0]{\ion{K}{i}\, IR}
\newcommand{\kirb}[0]{\ion{K}{i}\,IR$_{\rm blue}$}
\newcommand{\kirr}[0]{\ion{K}{i}\,IR$_{\rm red}$}
\newcommand{\kopt}[0]{\ion{K}{i}\, VIS}
\newcommand{\koptb}[0]{\ion{K}{i}\,VIS$_{\rm blue}$}
\newcommand{\koptr}[0]{\ion{K}{i}\,VIS$_{\rm red}$}

\usepackage{txfonts}

\usepackage[normalem]{ulem}

\begin{document}

\title{The CARMENES search for exoplanets around M dwarfs}
\subtitle{Diagnostic capabilities of strong \ion{K}{i} lines for photosphere and chromosphere\thanks{Full Table 2 is only available in 
electronic form
at the CDS via anonymous ftp to cdsarc.u-strasbg.fr (130.79.128.5)
or via http://cdsweb.u-strasbg.fr/cgi-bin/qcat?J/A+A/}}

\author{B. Fuhrmeister\inst{\ref{inst1}}, S. Czesla\inst{\ref{inst1}},  E. Nagel\inst{\ref{inst1},\ref{inst13}}
  \and   A.~Reiners\inst{\ref{inst2}}
  \and J. H. M. M. Schmitt\inst{\ref{inst1}}
  \and  S.~V.~Jeffers\inst{\ref{inst2}}
  \and  J.~A.~Caballero\inst{\ref{inst3}}
  \and  D.~Shulyak\inst{\ref{inst2},\ref{inst6}}
  \and  E.~N.~Johnson\inst{\ref{inst2},\ref{inst15}} 
  \and M.~Zechmeister\inst{\ref{inst2}}
  \and  D.~Montes\inst{\ref{inst10}}
  \and  \'A.~L\'opez-Gallifa \inst{\ref{inst10}}
  \and   I.~Ribas\inst{\ref{inst4},\ref{inst5}}
  \and  A.~Quirrenbach\inst{\ref{inst7}}
  \and P.~J.~Amado\inst{\ref{inst6}}
  \and  D.~Galad\'{\i}-Enr\'{\i}quez\inst{\ref{inst12}}
  \and  A.~P.~Hatzes\inst{\ref{inst13}}
  \and  M.~K\"urster\inst{\ref{inst14}}
  \and  C.~Danielski\inst{\ref{inst6}}
  \and  V.~J.~S.~B\'ejar\inst{\ref{inst8},\ref{inst9}}
  \and  A. Kaminski\inst{\ref{inst7}}
  \and  J.~C.~Morales\inst{\ref{inst5}}
  \and  M.~R.~Zapatero Osorio\inst{\ref{inst16}}}


\institute{Hamburger Sternwarte, Universit\"at Hamburg, Gojenbergsweg 112, D-21029 Hamburg, Germany\\
  \email{bfuhrmeister@hs.uni-hamburg.de}\label{inst1}
       \and
        Th\"uringer Landessternwarte Tautenburg, Sternwarte 5, D-07778 Tautenburg, Germany\label{inst13} 
        \and
        Institut f\"ur Astrophysik, Friedrich-Hund-Platz 1, D-37077 G\"ottingen, Germany\label{inst2} 
        \and
        Centro de Astrobiolog\'{\i}a (CSIC-INTA), ESAC, Camino Bajo del Castillo s/n, E-28692 Villanueva de la Ca\~nada, Madrid, Spain \label{inst3}
        \and
        Instituto de Astrof\'isica de Andaluc\'ia (CSIC), Glorieta de la Astronom\'ia s/n, E-18008 Granada, Spain\label{inst6} 
        \and
       Max-Planck-Institut f\"ur Sonnensystemforschung, Justus-von-Liebig-Weg 3,37077 G\"ottingen, Gemany\label{inst15}
        \and
        Facultad de Ciencias F\'{\i}sicas, Departamento de F\'{\i}sica de la Tierra y Astrof\'{\i}sica; IPARCOS-UCM (Instituto de F\'{\i}sica de Part\'{\i}culas y del Cosmos de la UCM), Universidad Complutense de Madrid, E-28040 Madrid, Spain\label{inst10} 
        \and
        Institut de Ci\`encies de l'Espai (CSIC), Campus UAB, c/ de Can Magrans s/n, E-08193 Bellaterra, Barcelona, Spain\label{inst4}
        \and
        Institut d'Estudis Espacials de Catalunya, E-08034 Barcelona, Spain\label{inst5}
           \and 
        Landessternwarte, Zentrum f\"ur Astronomie der Universit\"at Heidelberg, K\"onigstuhl 12, D-69117 Heidelberg, Germany\label{inst7} 
                \and
        Centro Astron\'omico Hispano-Alem\'an, Observatorio Astron\'omico de Calar Alto, Sierra de los Filabres, E-04550 G\'ergal, Almer\'{\i}a, Spain\label{inst12} 
                \and
        Max-Planck-Institut f\"ur Astronomie, K\"onigstuhl 17, D-69117 Heidelberg, Germany\label{inst14}
        \and
        Instituto de Astrof\'{\i}sica de Canarias, c/ V\'{\i}a L\'actea s/n, E-38205 La Laguna, Tenerife, Spain\label{inst8}
        \and
        Departamento de Astrof\'{\i}sica, Universidad de La Laguna, E-38206 Tenerife, Spain\label{inst9} 
       \and
       Centro de Astrobiolog\'{\i}a (CSIC-INTA), Carretera de Ajalvir, km 4. E-28850
Torrej\'{\o}n de Ardoz, Madrid, Spain\label{inst16}
       }
        
\date{Received dd/07/2021; accepted dd/mm/2021}

\abstract{There are several strong \ion{K}{i} lines found in the spectra of M dwarfs, among them
  the doublet near 7700\,\AA\, and another doublet near 12\,500\,\AA. We 
  study these optical and near-infrared  doublets in a sample of 324 M dwarfs,
  observed with CARMENES, the high-resolution optical and near-infrared spectrograph at Calar Alto, and
  investigate how well the lines can be used as photospheric and chromospheric diagnostics.
  Both doublets have a dominant photospheric component in inactive stars and can be used
  as tracers of effective temperature and gravity. For variability studies
  using the optical doublet, we
  concentrate on the red line component because this is less prone to artefacts from
  telluric correction in individual spectra. The optical doublet lines are sensitive to
  activity, especially for M dwarfs later than M5.0\,V where the lines develop an emission
  core. For earlier type M dwarfs, the red component of the optical doublet lines is also correlated
  with H$\alpha$ activity. We usually find positive correlation for stars with H$\alpha$ in emission,
  while early-type M stars with H$\alpha$ in absorption show anti-correlation. During flares,
  the optical doublet lines can exhibit strong fill-in or emission cores for our latest spectral types.
  On the other hand, the near-infrared doublet lines 
  very rarely show correlation or anti-correlation to H$\alpha$ and do not change line shape significantly
  even during the
  strongest observed flares. Nevertheless, the near-infrared doublet lines show notable resolved Zeeman splitting for about 20
  active stars which allows to estimate the magnetic fields $B$.   
  
      }
    
\titlerunning{Potassium lines}
\authorrunning{B. Fuhrmeister et~al.}
\maketitle


\section{Introduction}

The spectra of M dwarfs display several strong \ion{K}{i} lines. There is a
neutral potassium doublet  at 7667.009\,\AA\, and  7701.084\,\AA\,
(vacuum wavelengths), which are denoted as the \kopt\, lines here.  Moreover, in addition to
several weaker \ion{K}{i} lines, five strong
lines are found at significantly redder wavelengths: the near-infrared (NIR) doublet \kir\, 
lines at 12\,435.676\,\AA\, and 12\,525.591\,\AA\, and 
the  \ion{K}{i} doublet  at 11\,693.419\,\AA\, and 11\,776.060\,\AA\, near a companion line at 11\,772.859\,\AA.
These potassium lines are known to trace chromospheric activity, but also to be
sensitive to photospheric parameters. Furthermore, they are used as diagnostic lines in studies of exoplanet atmospheres
\citep[e.g.][]{Burrows2003, Sedaghati2016, Alexoudi2018, Oshagh2020}. 

Chromospheric activity for M dwarfs is most often traced by the H$\alpha$ line \citep{Gizis2000,Jeffers2018},
which is easily accessible for these stars.
With the advent of spectrographs also covering the NIR, interest
arose in new
activity-tracing lines in this wavelength range. While the chromospheric \ion{He}{i} line at 10830\,\AA\,
has been studied thoroughly \citep{Zirin1982, Sanz-Forcada2008, Andretta2017, hepaper},
other lines have received less attention. For example, \citet{Schmidt2012}
observed emission during large flares in the Paschen lines Pa$\beta$,
Pa$\gamma$, and the \ion{He}{i} line at 10\,830\,\AA,  but \citet{Robertson2016} could not find
a correlation
of the Pa$\delta$ line with H$\alpha$ for time-series observations of the M5.5\,V star
Proxima Centauri in quiescence. 
Nevertheless, these authors observed a strong correlation of a combined \kopt\ line index
with H$\alpha$ 
with a Spearman's rank correlation coefficient of 0.83 and $p< 10^{-45}$.
On the other hand, \citet{Kafka2006} found an anti-correlation of the absorption strength
of the redder optical \ion{K}{i} line with H$\alpha$
comparing stars in the Praesepe open cluster. These latter authors also found the anti-correlation to
be more pronounced for the group of early-type M dwarfs rather than for late-type M dwarfs.

The  \kopt\, lines have additionally been used as exoplanet
atmosphere diagnostic \citep[e.g.][]{Sing2011, Pont2013, Sedaghati2016, Alexoudi2018}. On the other hand,
\ion{K}{i} detection has been highly debated for some planets. For example, a $Hubble$/STIS
detection of potassium in the atmosphere of WASP-31b could not be validated by ground
based data \citep{Sing2015,Gibson2017,McGruder2020}. \citet{Oshagh2020} caution that
even broad band features with a possible origin in atmospheres of exoplanets can
actually be caused by stellar activity. Therefore, a good understanding of the activity
signatures expected in the \kopt\ lines is desirable, but this has never been
investigated in detail because of the strong telluric contamination of this region
by O$_{2}$ lines.
In a chromospheric activity study of M dwarfs, \citet{Patrick}, for example, exclude the \kopt\, 
lines from their analysis of chromospheric tracer lines,
which used spectra taken with the CARMENES spectrograph. Precise telluric correction for CARMENES
spectra has
meanwhile become available \citep{Evangelos}, and we intend to study not only the \kopt\, lines
but also the \kir\, lines.  

Besides use as chromospheric activity tracers, the potassium lines have a large
diagnostic potential for photospheric parameters of  M stars. The
photospheric component of the \ion{K}{i} lines is known to
undergo strong changes from M to L spectral types, which was described by
\citet{Kirkpatrick1999} and \citet{Kirkpatrick1991} for the  \ion{K}{i} lines doublet at 7700 \AA.
The general behaviour of these lines was also studied theoretically by
\citet{Pavlenko2000} for L dwarfs. The general behaviour of the \ion{K}{i}
lines in the optical and the near-infrared are the same. For example,
a combination of both lines of the \ion{K}{i} doublet
at 12435\,\AA\, and 12525\,\AA\, was studied by \citet{Gorlova2003}.
Later,  \citet{McLean} studied
all strong  \ion{K}{i} lines in the $J$-band:  
The two \kir\, lines at 11\,693\,\AA\, and 11\,776\,\AA\,
have approximately the same strength, while the line at 11\,772\,\AA\, is weaker. 
For the \kir\, doublet at 12435\,\AA\, and 12525\,\AA, the redder line is always deeper \citep{McLean}.
Generally, the \ion{K}{i} lines in the optical as well as in the IR
broaden to later spectral types. While the lines display nearly no wings in M2.0 dwarfs,
wings develop through the M spectral sequence and
L dwarfs have extended wings. \citet{McLean} found that the line depth and
pseudo equivalent width (pEW)
also grow along the M type sequence and peak at around mid-L. In contrast, the half width
grows until mid-T dwarfs which marked the end of the spectral sequence observed by 
\citet{McLean}.
The growth of the line width for stars later than M7.0\,V can be explained as follows: 
with decreasing temperatures, the lines should weaken because the transition levels become 
less populated. However, as dust grains settle more and more below the photosphere, with 
decreasing temperature the transparency improves. Hence, line formation is observed at 
much greater depths and pressures \citep{Saumon2003}. 
With increasing pressure the lines develop broad wings caused by collisional broadening,
primarily with H$_{2}$ molecules \citep[e.\,g.][]{Burrows2003}.

Here we study the strong potassium lines \kopt\ and \kir\ along the M dwarf sequence
and their diagnostic potential regarding photosphere and magnetic chromospheric activity. First,
we deal with the photospheric parts of the lines and their use as possible
stellar parameter indicators. Second, we  inspect how activity
manifests itself in these lines and if they are sensitive to Zeeman broadening.

Due to the fact that, in M dwarfs, Zeeman splitting often only leads to additional line broadening,
radiative transfer models are needed to model the spectra correctly.
Nonetheless, magnetic fields have been measured in M stars.
For example, \citet{Johns-Krull1996} measured the $\sigma$ components of Zeeman splitting in
the wings of a \ion{Fe}{i} line at 8468.40\,\AA\, in several M dwarfs
and was able to detect strong magnetic fields of 3.8 and 2.6\,kG in EV~Lac and Gliese~729,
respectively, with filling factors $f$ of 50\,\%.
\citet{Reiners2006} used the FeH lines at about 1 $\mu$m to classify the magnetic
field for 20 M2 -- M9 dwarfs as weak, intermediate, or strong. For instance, they
report $Bf$ > 3.9\,kG for YZ~CMi, 
$Bf$=2.9\,kG for AD~Leo, and $Bf$=2.3\,kG for
vB~8. $Bf$ here is the product of the magnetic field $B$ and
the filling factor $f$. \citet{Shulyak2019} made use of a direct magnetic
spectrum
synthesis to improve this kind of measurement
and derived magnetic fields in a sample of 29 M dwarfs using
CARMENES spectra.

The paper is structured as follows: In Sect. \ref{sec:obs} we present the data, their reduction,
and the method for the pseudo-equivalent width measurement. In Sect. \ref{sec:results} we
deal first with the usability of the blue optical line in Sect. 
\ref{sec:tellcorr}, then with the usage
 of the potassium lines for determination of photospheric parameters
in Sect. \ref{sec:photosphere}, followed by the diagnostic capabilities for stellar activity
in Sect. \ref{sec:activity}. There we concentrate on the activity characterisation by correlation
with H$\alpha$ in Sect. \ref{Sec:corr}, the behaviour during strong flares in Sect. \ref{emisscores},
and the detection of resolved Zeeman splitting in some stars in Sect. \ref{Zeeman}. We
present concluding remarks in Sect. \ref{conclusion}.

\section{Observations and measurement method}
\label{sec:obs}
\subsection{Available data and their reduction}

All spectra used for the present analysis were taken
with the CARMENES spectrograph installed at the 3.5\,m Calar Alto 
telescope \citep{CARMENES1}.
CARMENES is a spectrograph covering the wavelength range
from 5200 to 9600\,\AA\, in the visual channel (VIS) and from 9600 to 17\,100\,\AA\, in
the NIR
channel. The instrument provides a spectral resolution of
$\sim$ 94\,600 in VIS and $\sim$ 80\,400 in NIR. 
The CARMENES consortium devoted 750~nights to conducting a 
survey of $\sim$350 M~dwarfs to find low-mass exoplanets \citep{AF15a, Reiners2017}.
As the cadence of the spectra is optimised for the planet search, usually
no continuous time-series are obtained.

In our analysis, we consider a
sample of 324 M~dwarfs, excluding known close binaries \citep{Baroch2018,Schweitzer2019},
resulting in a study of 
about 15\,000 spectra taken before July 2020. The stellar spectra
were reduced using the CARMENES reduction pipeline
\citep{pipeline,Caballero2}. Subsequently, we corrected them for barycentric and radial velocity
motions.
We carried out a correction for telluric absorption lines \citep{Evangelos} 
using the {\tt molecfit}
package\footnote{\tt{https://www.eso.org/sci/software/pipelines/skytools /molecfit}}, but
did not correct for airglow emission lines.

\subsection{Measurement of equivalent width }

We measured the pEW of the blue \kopt\, line at 7667.009\,\AA\, (hereafter \koptb) and of the red \kopt\ line at 7701.084\,\AA\, (hereafter \koptr),
the blue \kir\, doublet line at 12435.676\,\AA\, (hereafter \kirb) and the
red \kir\ line at 12525.567\,\AA\, (hereafter \kirr), and 
the H$\alpha$ line for comparison purposes. 
The integration ranges and the reference bands for these lines are given in Table~\ref{ew}. The pEW is calculated using
\begin{equation}
        \mathrm{pEW}=\int (1- F_{core}/F_{0})d\lambda
,\end{equation}  
where $F_{core}$ is the flux density in the line band, $F_{0}$ is the mean flux density in the two reference intervals (representing the pseudo continuum), and $\lambda$ denotes wavelength.
The H$\alpha$ line allows us to discriminate between active and inactive stars:
 we define H$\alpha$ in absorption by pEW(H$\alpha$) $> -0.6$\,\AA\, 
 \citet{hepaper} and call these inactive.

We exclude three strong \ion{K}{i} lines in the NIR at
11\,693.419\,\AA, 11\,772.859\,\AA, and 11\,776.060\,\AA\, from our analysis because they
are contaminated by OH airglow lines at 11\,696.44\,\AA\, and 11\,771.00\,\AA\, \citep{Oliva2015}, which
are not corrected
by the method of \citet{Evangelos}. Another \kir\ line at 11\,693.419\,\AA\, is heavily
blended with an \ion{Fe}{i} line at 11\,693.176\,\AA, which can be nearly as strong as the
\ion{K}{i} line. Finally, the two lines at 11\,772.859\,\AA\, and 
11\,776.060\,\AA\, are increasingly blended
with each other toward
the mid- and late M~dwarf regime.

To avoid the region heavily contaminated by telluric features bluewards of the 
\koptb\, line, we choose the same reference region for both doublet lines with
one reference region between the two lines and one redwards of the 
\koptr\, line. As photospheric \ion{K}{i} lines develop
broad wings, we caution that, for stars later than M4.5\,V, the reference bands
are situated in the far wings of the lines for \kopt. We used narrow wavelength
ranges to compute the pEW only for the line centre, which is most sensitive
to activity-related `fill-in' or emission in the line core. The
difference between fill-in and emission cores can be best seen in Figs. \ref{timeseries} 
(bottom panel) and \ref{flares} (left panels).
Therefore, our integration band is optimised for
detecting chromospheric variability, which would be diluted in a setup with broader
integration bands optimised for stellar parameter determination.
Also, for fast rotators with \mbox{\vsini\, $>$ 15\,km\,s$^{-1}$}, all pEWs
are underestimated compared to pEWs for the whole line; see Sect. \ref{sec:photosphere}. We also caution that our pEW is dependent on the resolution of 
the spectra. In particular, lower resolution is expected to carry photospheric signal
into the integration band while narrow chromospheric emission
would be blurred out of it. The reverse occurs at higher resolution until
the instrumental resolution surpasses the width of the natural spectral
features and
higher resolution ceases to noticeably affect the spectrum. Because the
reference bands are broad, resolution effects are much weaker there. While
the details
depend on the exact spectral shape, we expect negligible changes in the pEWs
as long as the central integration band covers a minimum of three resolution
elements. This translates into
a minimum resolution of 50 000 for the \kopt\ lines and 77 000 for the \kir\
lines required to reproduce the pEWs with our setup. Yet, lower spectral
resolution
and potentially wider bands may of course still yield comparable if not
identical results.

Thus, time-series of pEW measurements were obtained for each star. 
From these pEW time-series, we calculated the median pEW
and the median average deviation about the median (MAD), which is 
a robust estimator of the scatter \citep[][]{Hampel1974, Rousseeuw1993, Czesla2018}.
The uncertainty of individual
pEW measurements strongly depends on the S/N of the spectra, which may be an issue 
for H$\alpha$ and the \kopt\ lines. Typically, the S/N is better
for bright, early M~dwarfs than for fainter mid-type M dwarfs because the maximum
 exposure time
in the survey was 1800\,s. We did not compute pEWs when the S/N in the pseudo-continuum was
lower than 10, and therefore pEW measurements of the H$\alpha$
line are missing for some of the latest M~dwarfs.
\begin{table*}
\caption{\label{ew} Parameters of the pEW calculation. }
\footnotesize
\begin{tabular}[h!]{lcccc}
\hline
\hline
\noalign{\smallskip}

Line           & wavelength   & Width  & Reference band 1 & Reference band 2 \\
& [\AA] & [\AA] & [\AA] & [\AA]\\
\noalign{\smallskip}
\hline
\noalign{\smallskip}
H$\alpha$& 6564.60 & 1.60 & 6537.43--6547.92 & 6577.88--6586.37 \\
\koptb    & 7667.01 & 0.50 & 7687.00--7689.00 & 7703.00--7705.50 \\
\koptr    & 7701.08 & 0.50 & 7687.00--7689.00 & 7703.00--7705.50 \\
\kirb     & 12\,435.68  & 0.50 & 12\,369.00--12\,373.50& 12\,531.00--12\,533.00\\
\kirr     & 12\,525.57  & 0.50 & 12\,369.00--12\,373.50 & 12\,531.00--12\,533.00\\
\noalign{\smallskip}
\hline

\end{tabular}
\normalsize
\end{table*}

\subsection{Stellar parameters}

\begin{figure}[h!]
\begin{center}
\includegraphics[width=0.5\textwidth, clip]{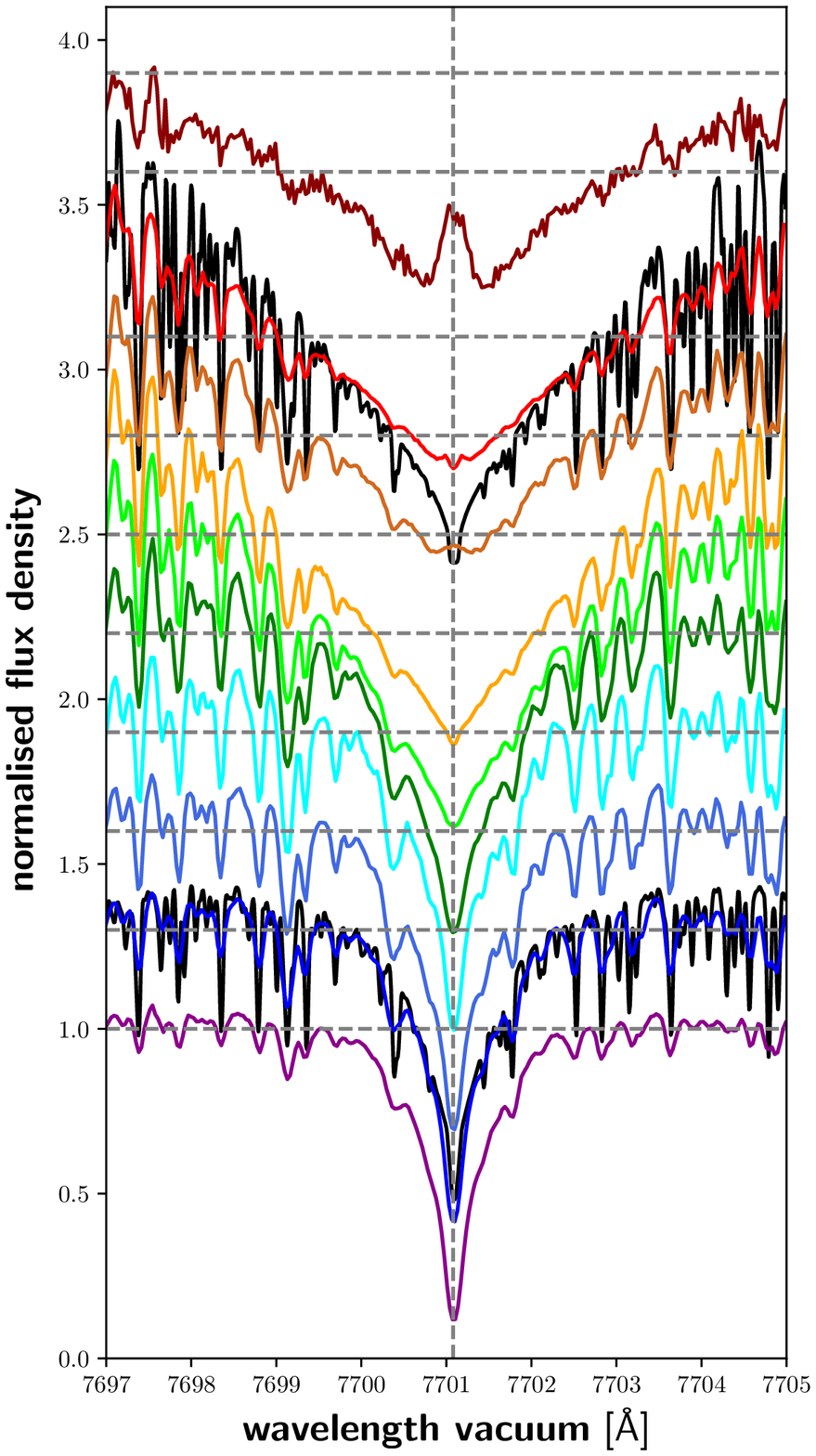}\\
\caption{\label{kopttemp1} Spectral subtype sequence for the wavelength region around \koptr.
  There is an offset added to the normalised spectra for convenience, which is larger
  for the M7.0 spectrum. For comparison we show PHOENIX models in black for two of the
  stars; more details are given in the main text.
  The shown spectra correspond to the following stars from bottom to top (Karmn number
        and corresponding line colour in parenthesis):
  M0.0\,V: HD~23453 (J03463+262, purple);
  M1.0\,V: GJ~2 (J00051+457, blue);
  M2.0\,V: GJ~47 (J01013+613, dark blue);
  M3.0\,V: G~244-047 (J02015+637, cyan);
  M4.0\,V: LP~768-113 (J01339--176, green);
  M4.5\,V: YZ~Cet (J01125--169, lime green);
  M5.5\,V: GJ~1002 (J00067--075, orange);
  M6.0\,V: CN~Leo (J10564+070, brown);
  M7.0\,V: Teegarden's Star (J02530+168, red);
  M8.0\,V: vB~10 (J19169+051S, dark red)
}
\end{center}
\end{figure}

In the following, we compare our measured pEWs to effective temperatures,
$T_{\rm eff}$, surface gravity,
$\log\,g,$ and metallicity [Fe/H]. As stellar parameters
already exist for all our sample stars, we refrained from computing these by ourselves and instead used
the ones reported by \citet{Schweitzer2019}, who employed the method of
\citet{Vera}.
These two studies used PHOENIX
photospheric models published by \citet{Husser} to derive the stellar parameters
based on fits to CARMENES spectra. 
We show two of their best-fit PHOENIX models in Fig.~\ref{kopttemp1}, which demonstrates
the good fit of the models in the wings. The line core, however, can be influenced by chromospheric
activity, which is not taken into account in the (purely photospheric) spectral models.

\section{Results}\label{sec:results}

\subsection{Usability of the \koptb\ line}
\label{sec:tellcorr}

\begin{figure}
\begin{center}
\includegraphics[width=0.5\textwidth, clip]{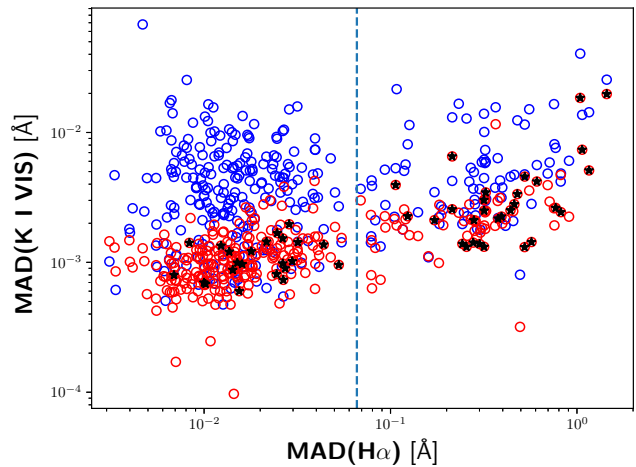}\\
\caption{\label{tellcor}  MAD(\koptr) (red open circles) and MAD(\koptb) (blue open circles) as a function of
    MAD(H$\alpha$). Black symbols denote the dwarfs with a correlation between  
        pEW(\koptr) and pEW(H$\alpha$) as defined in Sect. \ref{Sec:corr}. 
        The dashed vertical line  
        marks the
        gap between weakly and strongly variable stars, which corresponds roughly to
        inactive and active stars.
}
\end{center}
\end{figure}

For the \kopt\, doublet, the blue line is much more prone to telluric line contamination
than the red line, which provides an opportunity to test the telluric correction.
To that end, we compare relative variability observed in the line cores as measured by
the variation in pEW(\koptb) and pEW(\koptr). In particular,
we investigate the respective MADs, which we show as a function of MAD(H$\alpha$)
in Fig.~\ref{tellcor}.
The MAD of pEW(\koptb) is typically higher than that for the red line. Such high variations in any of the \kopt\,
lines
are not expected, particularly for the inactive stars
with little variation in H$\alpha$, . 

Several factors can contribute to the total scatter observed in the lines.
First, we consider statistical noise with variance, $\sigma_{s,x}^2$, where $x$ denotes
either $r$ or $b$ to identify the red and blue \ion{K}{i} lines. This term represents measurement
noise and instrumental effects. Second, activity contributes to the scatter of pEW  
measurements, which we name $\sigma^2_{a,x}$. Third, inaccuracies in the telluric correction can add
to the observed scatter, which we summarise in the term $\sigma^2_{t,x}$. 
Assuming that these terms are independent and approximately Gaussian,
their variances can be added together, so that we obtain
\begin{equation}
    \sigma^2_{x} = \sigma^2_{s,x} + \sigma^2_{a,x} + \sigma^2_{t,x} \; .
\end{equation}
The statistical
variances are equal for both lines as can be seen from the very similar statistical errors for the flux density in both lines. Similarly, the emission oscillator strengths or, equivalently,
Einstein coefficients for spontaneous emission are almost identical. As the lines both form in the same chromospheric regions, this implies
that activity-induced emission for both lines is also very similar,  and so the activity-related
variances can also be assumed to be equal. Given that the red line suffers from marginal telluric
contamination, we assume $\sigma^2_{t,red} = 0$ and it follows that the observed excess scatter
in pEW(\koptb) is attributable to the telluric variance term, $\sigma^2_{t,blue}$.
This is consistent with
the correlation between pEW(\koptb) and pEW(\koptr) tending to be weak for the time-series of the stars with excess scatter in pEW(\koptb).
We therefore only use the \koptr\ line in our further analysis.

\subsection{Line profile development with spectral type and gravity}\label{sec:photosphere}

As predicted by synthetic spectra,
we observe a strong dependence of pEW(\kopt) and pEW(\kir) on $T_{\rm eff}$.
In Fig.~\ref{kopttemp1}, we show a spectral sequence of line profiles of the 
\koptr\, line, which  shows
the development of broad wings towards later spectral types.
Moreover, excess emission may be distinguished in the lines cores
in subtypes of M5.5\,V\, stars and later, which is discussed in more detail in Sect.~\ref{emisscores}.
The \kir\, lines also broaden toward later spectral
types, but no systematic appearance of emission cores for the latest spectral
types is observed. However, some stars show line shapes reminiscent of chromospheric
emission cores, which is discussed in more detail in Sect.~\ref{Zeeman}.

\begin{figure*}
\begin{center}
\includegraphics[width=0.5\textwidth, clip]{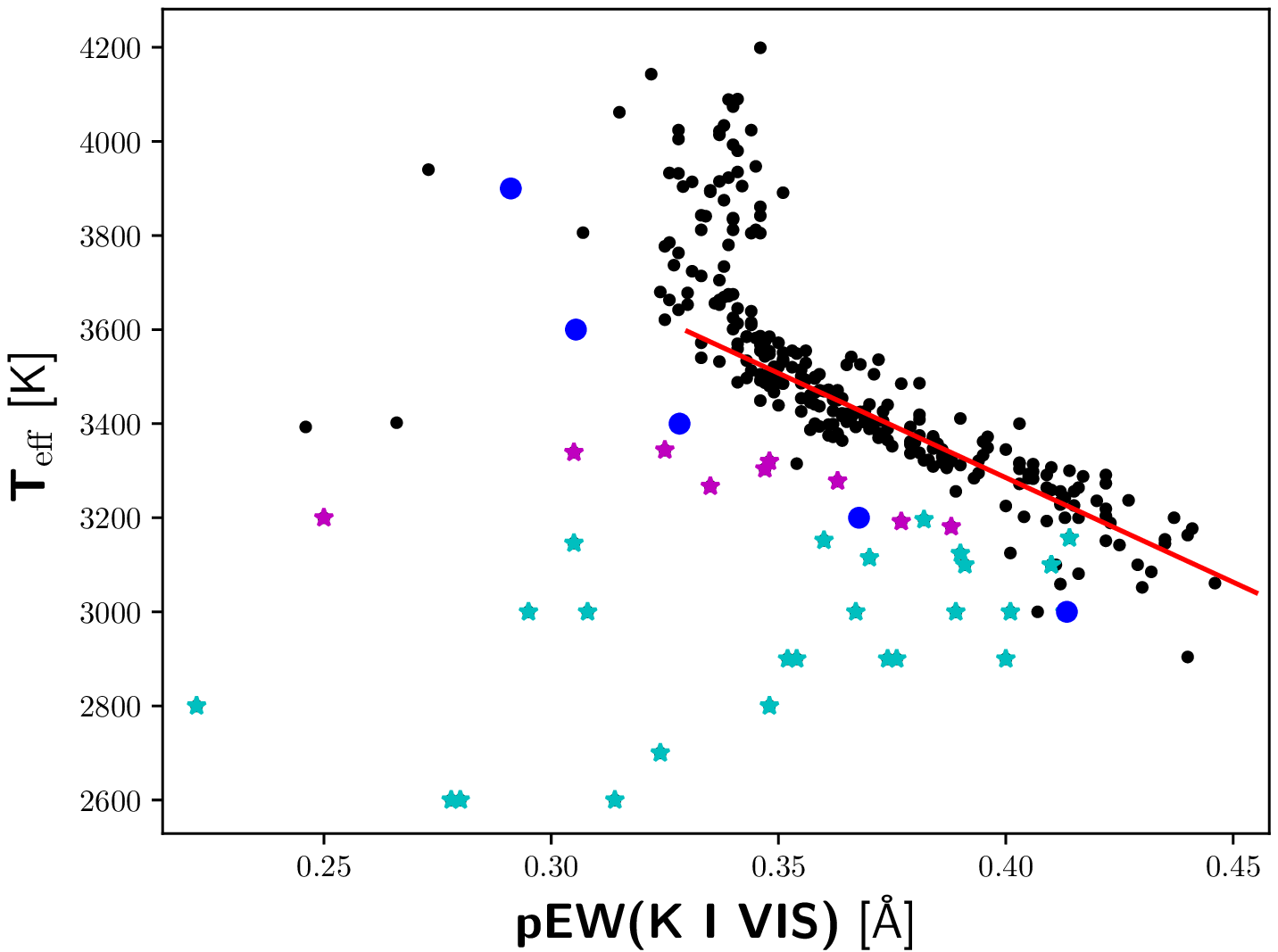}
\includegraphics[width=0.5\textwidth, clip]{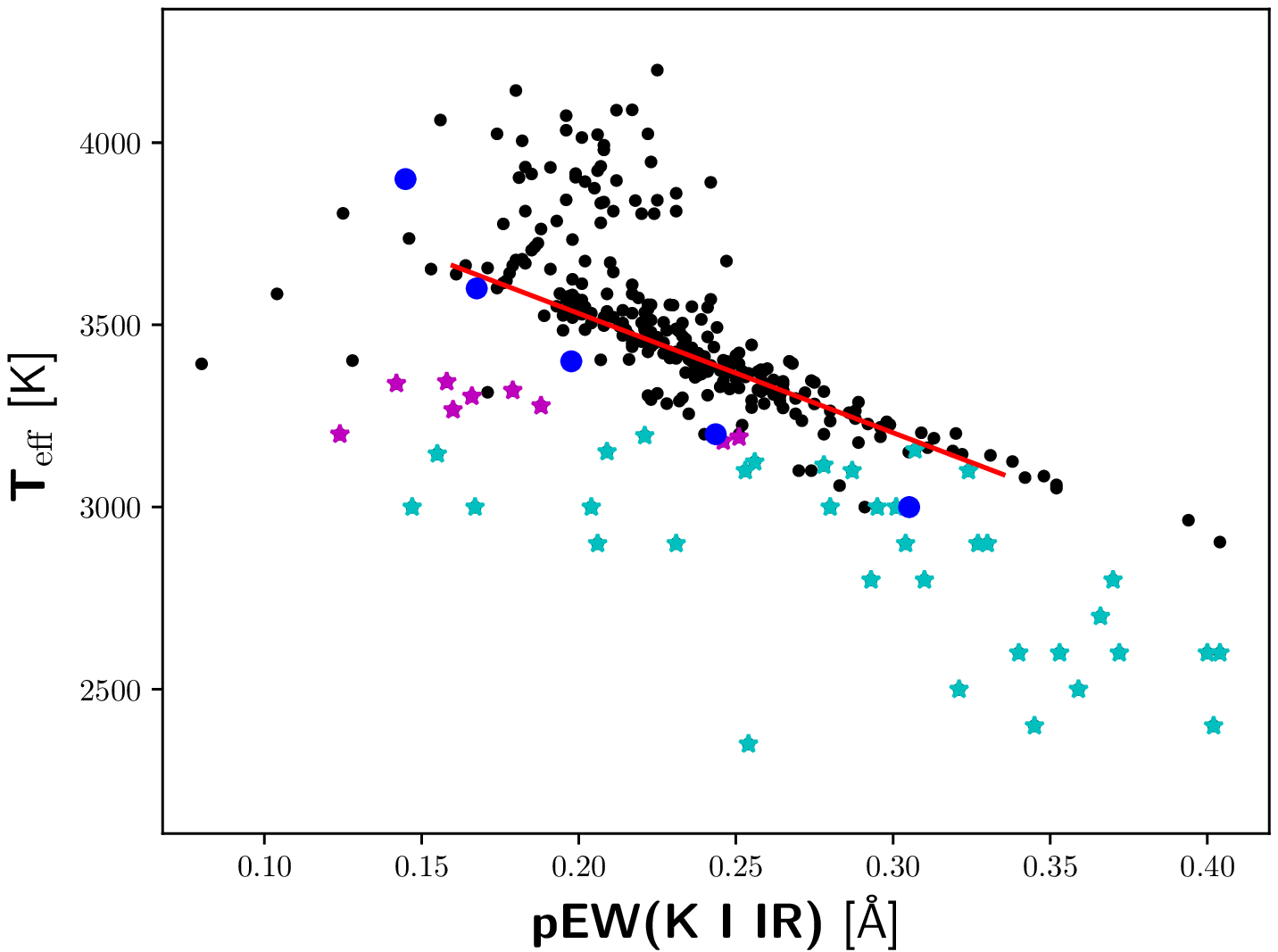}\\
\caption{\label{kopttemp}  Effective temperature $T_{\mathrm{eff}}$ as a
  function of pEW(\koptr) \emph{(left)} and pEW(\kirr) \emph{(right)}.
  Stars marked in magenta are fast rotators
  with \vsini\, $>$ 15\,km\,s$^{-1}$. Stars marked in cyan have spectral type M5.0\,V or later
  and have measurable \vsini\, $>$ 2\,km\,s$^{-1}$. Blue circles mark the
  pEW as calculated from PHOENIX spectra taken from \citet{Husser}.
}
\end{center}
\end{figure*}

In Fig.~\ref{kopttemp}, we show a comparison between pEW(\kopt) and 
$T_{\rm eff}$ (as determined by \citet{Schweitzer2019}). A close linear relation
is apparent for all but the earliest M~dwarfs with $T_{\rm eff} \ge 3600$\,K.
For comparison we also show pEW(\koptr) calculated from PHOENIX spectra
from the library by \citet{Husser}. The deviation between the PHOENIX
spectra and observations in the line core may be attributed to two factors: 
First, the photospheric
models are calculated using local thermal equilibrium (LTE), which may lead to
small deviations in the core compared to non-LTE models. This may apply for the smaller
deviation in the \kir\ lines. Second, the models do not include any activity.
As the \kir\ lines react less to activity as detailed in Sect. 
\ref{Sec:corr}, activity may account for the larger systematic deviation between PHOENIX
and observed spectra in the \koptr\
line. As there seems to be a shift of about 200 K between models and observations,
this may indicate that the \koptr\ line has a chromospheric contribution even for inactive stars and is formed in the lower chromosphere.

Observed outliers are due to several factors: (i) Fast rotation.
We mark stars with 
\mbox{\vsini\, $>$ 15\,km\,s$^{-1}$}. Fast rotation flattens the line profiles and causes the lower
pEW measurements as we focus only on the line centre. Many of these objects
are also young (age below 100 Ma), and therefore subject to low gravity, which
may also decrease their pEW. (ii) Chromospheric activity.
We also mark stars of type M5.0\,V and later with a
measured \vsini\, of 2\,km\,s$^{-1}$ or more. While chromospheric emission cores
are noted only at M5.5\,V and later, some chromospheric contribution sets in at M5.0\,V
already.
(iii) Miscorrections of telluric lines or spectral misalignment.
This explains in particular all non-marked outliers with
pEW(\kopt) values lower than 0.3\,\AA.
Therefore, we exclude pEW(\kopt) below this threshold from the following considerations; also we exclude all outliers caused
by fast rotation or chromospheric activity.
In this case, the relation between $T_{\rm eff}$ and pEW(\kopt) for the 194 stars fulfilling also
3000\,K $<T_{\rm eff}< $3600\,K
leads to a Pearson's correlation coefficient of $-0.91$,
formally corresponding to a $p$-value of $10^{-78}$. The relation can be described by a
linear regression of the form
\begin{equation}
    T_{\mathrm{eff, \ion{K}{i}\,VIS\,red}}= -4442\,\mbox{K}\,\AA^{-1} \times \mathrm{pEW(\ion{K}{i}\,VIS_{red})} + 5062\,\mbox{K}
,\end{equation}
leading to a standard deviation of $T_{\rm eff, \ion{K}{i}\,VIS\,red}-T_{\rm Schw19}$ of 55\,K.

A similar trend is observed for the pEW(\kirr),
but again for $T_{\rm eff}>$  3600\,K  a
reversal in pEW(\kirr) is observed as shown in Fig. \ref{kopttemp}. Applying
again a linear regression leads to the relation
\begin{equation}
        T_{\mathrm{eff, \ion{K}{i}\,IR\,red}}= -3273\,\mbox{K}\,\AA^{-1} \times \mathrm{pEW(\ion{K}{i}\,IR_{red})} + 4186\,\mbox{K}  
,\end{equation}
which yields a standard deviation of 64\,K for $T_{\rm eff,\ion{K}{i}\,IR\,red}-T_{\mathrm{Schw19}}$.
The value of Pearson's correlation coefficient is $-0.88$ with a $p$-value of $10^{-66}$.

\begin{figure*}
\begin{center}
\includegraphics[width=0.5\textwidth, clip]{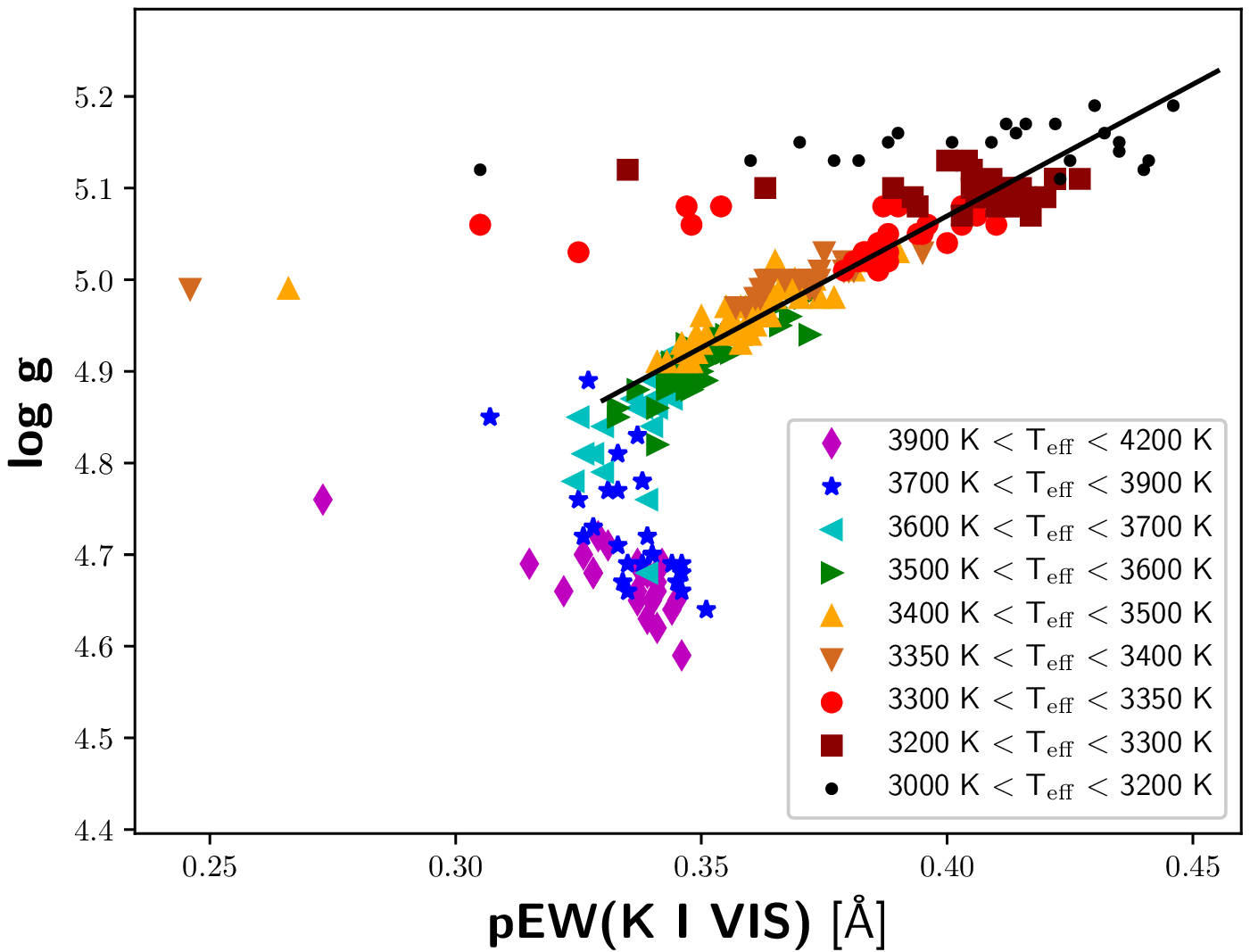}
\includegraphics[width=0.5\textwidth, clip]{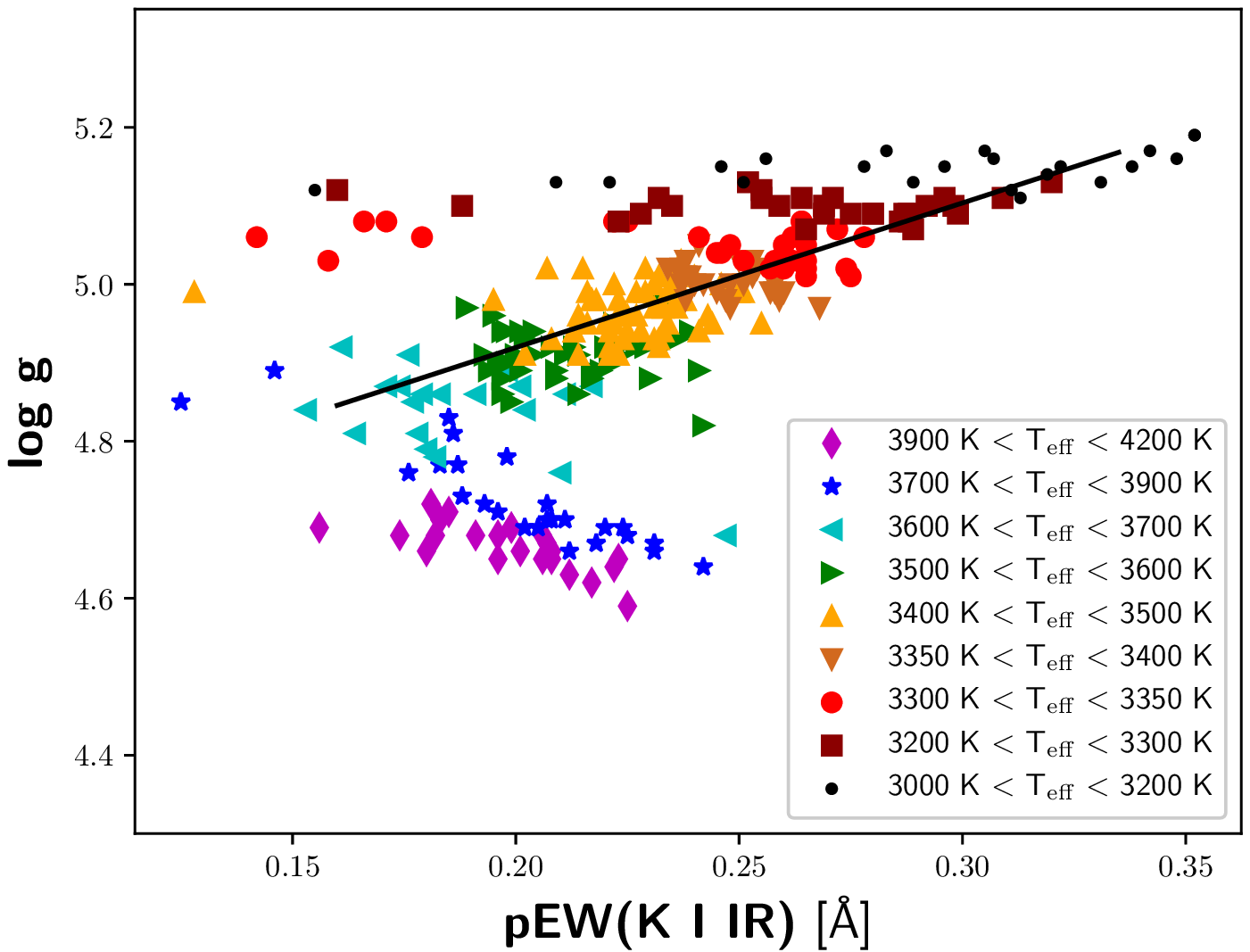}\\
\caption{\label{koptlogg}  $\log\,g$ as a function of pEW(\koptr) \emph{(left)}
  and pEW(\kirr) \emph{(right)}, respectively, 
  and effective temperature colour coded for both cases.}
\end{center}
\end{figure*}

If the effective temperature, $T_{\mathrm{eff}}$, is replaced by
$\log\,g$ as the dependent variable, a similar relation is found
for the slowly rotating main sequence dwarfs. For stars with 3600\,K $> T_{\mathrm{eff}} > ~$3000\,K 
this can be seen in Fig.~\ref{koptlogg}  with a reversal for stars with  
higher temperatures. This similarity in the  behaviour is expected, because
the effective temperature
and surface gravity are not independent variables for main sequence stars
as detailed for example in \citet{Schweitzer2019}.

The value of the Pearson's correlation coefficient between 
pEW(\koptr) and $\log\,g$ is 0.95 with a $p$-value of $10^{-102}$.
The best-fit relation is given by
\begin{equation}
        \log g_{\mathrm{\ion{K}{i}\,VIS\,red}}= 2.874\,\AA^{-1} \times \mathrm{pEW(\ion{K}{i}\,VIS_{red})} + 3.920\,{\rm dex}
,\end{equation}
with std($\log g_{\rm \ion{K}{i}\,VIS\,red}$-$\log g_{\rm Schw19}$) = 0.06\,dex.
For pEW(\kir), a somewhat less pronounced correlation is found with
Pearson's correlation coefficient yielding 0.81 and a $p$-value of $10^{-46}$.
The best-fit relation reads
\begin{equation}
        \log g_{\mathrm{\ion{K}{i}\,IR\,red}}= 1.785\,\AA^{-1} \times \mathrm{pEW(\ion{K}{i}\,IR_{red})} + 4.566\,{\rm dex,}
\end{equation}
and std($\log\,g_{\rm \ion{K}{i}\,IR\,red}$-$\log\,g_{\rm Schw19}$) = 0.09\,dex.

Finally, we compared pEW(\kopt) and pEW(\kir) to the metallicity [Fe/H] also
determined by \citet{Schweitzer2019}, but
could not identify any relation similar to the ones presented above.
We show our comparison in Fig. \ref{knirfeh} of the Appendix. The
Pearson's correlation coefficient between [Fe/H] and pEW(\kopt) is \mbox{--0.06}
with a $p$-value of 0.40, and for pEW(\kir) it is 0.24 with a $p$-value of
0.0004. Though the latter $p$-value formally indicates a correlation, it is much weaker
than for $T_{\rm eff}$ or $\log\,g$. Also computing a linear regression as above 
leads to a standard error of 0.17\,dex in the thus-determined [Fe/H]. As
the [Fe/H] values in our sample (again not considering the outliers) range only from
\mbox{--0.34}\,dex to 0.17\,dex, we do not consider a formal fit to be useful.

Therefore, in summary, a comparison with the results by \citet{Schweitzer2019} demonstrates that
for effective temperatures in the range $3000 < \mbox{$T_{\mathrm{eff}}$} < 3600$\,K,
the \kopt\ and \kir\ lines alone allow a good estimate of temperature and 
$\log\,g$, while an estimation of the metallicity is not possible with the
data.

\subsection{\kopt\, and \kir\, lines as activity tracers}\label{sec:activity}

The values of pEW(\kopt) cannot be used as a direct activity indicator because they
strongly depend on spectral type and are dominated by photospheric contributions.
This combination prevents comparison between stars of different spectral type and also makes
subtraction of a  quiescent template challenging.
As an illustration, we show the dependence of 
pEW(\kopt) on pEW(H$\alpha$) for each spectral type in Fig.~\ref{kopthalpha}.

As shown in Fig.~\ref{tellcor} and discussed further in Sect. \ref{Sec:corr},
the values of MAD(pEW(\koptr)) are correlated with MAD(pEW(H$\alpha$)).
In particular, we find a value of $0.6$ for the Pearson's correlation coefficient
and a $p$-value of $10^{-31}$, suggesting
a chromospheric origin of the variations. 
We do not consider pEW(\koptb) because of the remaining telluric contamination for many stars.
Moreover, when considering MAD(pEW(\kir)), we do not find a correlation to
MAD(pEW(H$\alpha$)), showing the relative insensitiveness of the \kir\ lines
to stellar activity. We show our findings for MAD(pEW(\kir)) in Fig. \ref{tellcorrir}. 
\begin{sidewaystable}
\caption{\label{measurements} Measured median pEWs, MADs, and correlation coefficients of the considered lines.$^{a }$}
\footnotesize
\begin{tabular}[h!]{llrrrrrrrrrrrr}
\hline
\hline
\noalign{\smallskip}
        Karmn&  Name       &Corr & p-val& MAD & MAD  &MAD & MAD & MAD& Median & Median& Median& Median& Median\\
        &    &     \koptr-      &      & (\koptb)&(\koptr) &(\kirb)&(\kirr) &(H$\alpha$)& pEW&   pEW & pEW& pEW&pEW \\
         &   & H$\alpha$  &  &  &  &  & & &(\koptb)&(\koptr) &(\kirb)&(\kirr) &(H$\alpha$)\\
          &  &        &   &    [\AA]       &  [\AA]  &  [\AA] & [\AA] & [\AA]& [\AA]  & [\AA] & [\AA] & [\AA] & [\AA]\\
\hline
        J00051+457 &    GJ 2  &   --0.089  &     0.541  &     0.002    &   0.001  &     0.001  &     0.001&       0.018  &     0.367   &    0.340&        0.164  &     0.202   &      0.347\\
        J00067-075 &GJ 1002   &   0.322  &     0.004  &     0.004   &    0.003  &     0.001  &     0.001  &     0.069  &     0.454  &     0.446   &     0.329  &     0.352   &     --0.057\\
        J00162+198E & LP 404-062          &   0.302  &     0.294   &    0.002   &    0.001   &    0.002  &     0.002&       0.017  &     0.407   &    0.383    &    0.229   &    0.259   &      0.139\\
        J00183+440 & GX And  &   --0.081   &    0.262    &   0.017    &   0.001  &     0.001  &     0.002  &     0.007   &    0.380   &    0.344   &     0.137   &    0.176   &      0.320\\
        J00184+440 & GQ And  &    0.324   &    0.000    &   0.003   &    0.001  &     0.001   &    0.001&       0.010 &      0.442    &   0.422    &    0.232  &     0.264     &    0.160\\
        J00286-066 & GJ 1012     &    --0.242   &    0.101  &     0.004  &     0.001  &     0.001  &     0.002  &     0.009   &    0.388   &    0.361   &     0.206   &    0.238   &      0.168\\
        J00389+306 & Wolf 1056  &   --0.019   &    0.927  &     0.003  &     0.002  &     0.001    &   0.001&       0.013  &     0.371   &    0.351   &     0.174   &    0.209   &      0.294\\
        J00570+450 & G 172-030    &   --0.660    &   0.020  &     0.006   &    0.001 &      0.001   &    0.001 &       0.028  &     0.391  &     0.369   &     0.197   &    0.229   &      0.154\\
        J01013+613 & GJ 47  &   --0.477   &    0.194  &     0.009  &     0.001  &     0.001 &      0.001 &      0.039   &    0.388   &    0.356   &     0.165   &    0.201  &       0.246\\
        J01019+541 & G 218-020    &    0.964   &    0.000  &     0.003  &     0.002  &     0.002   &    0.002 &       0.394 &      0.366    &   0.352   &     0.194  &     0.230    &    -4.136\\

\multicolumn{13}{c}{\ldots}\\ \hline
\end{tabular}

$^{a}$ The full table is provided at the CDS. We show here the first ten rows as  guidance.
\normalsize
\end{sidewaystable}

\subsubsection{Correlation with the H$\alpha$ line}\label{Sec:corr}

Making use of our time-series measurements, we  search for possible (anti-)correlation
of pEW(\koptr) and pEW(\kir) with pEW(H$\alpha$) in individual stars.
The relevant measurements are listed in Table~\ref{measurements} with the full table
available at CDS. In particular, the table gives a common name and also the CARMENES identification, Karmn, of each 
star and the value of the Pearson's correlation coefficient between pEW(\kopt)  and 
pEW(H$\alpha$) along with the corresponding $p$-value. Moreover,  
Table~\ref{measurements}
lists the median(pEW) and the MAD for H$\alpha,$  the red and blue  \kopt\ lines, and the red and blue \kir\ lines.

\begin{figure}[h!t]
\begin{center}
\includegraphics[width=0.5\textwidth, clip]{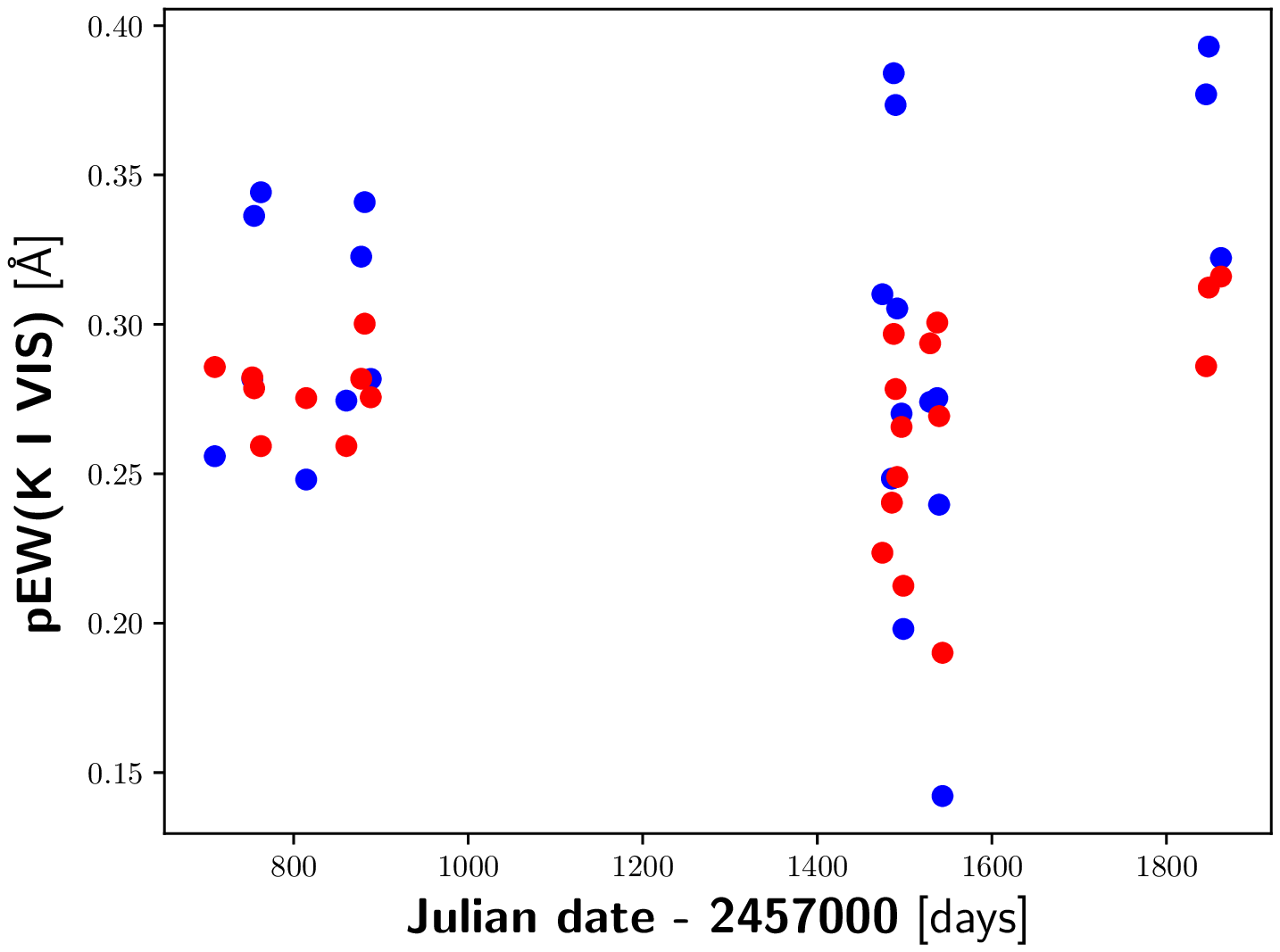}\\
\includegraphics[width=0.5\textwidth, clip]{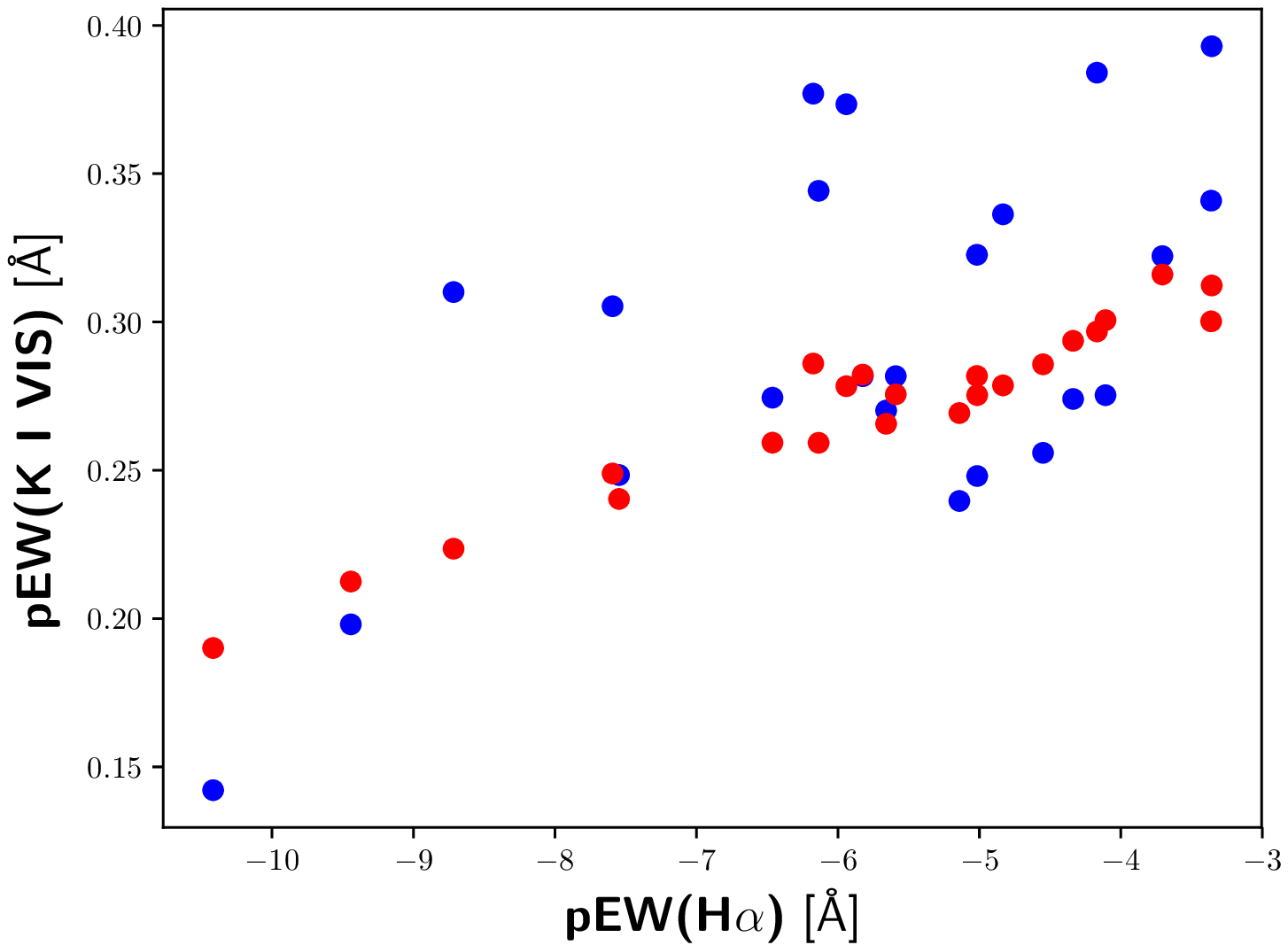}\\
\includegraphics[width=0.5\textwidth, clip]{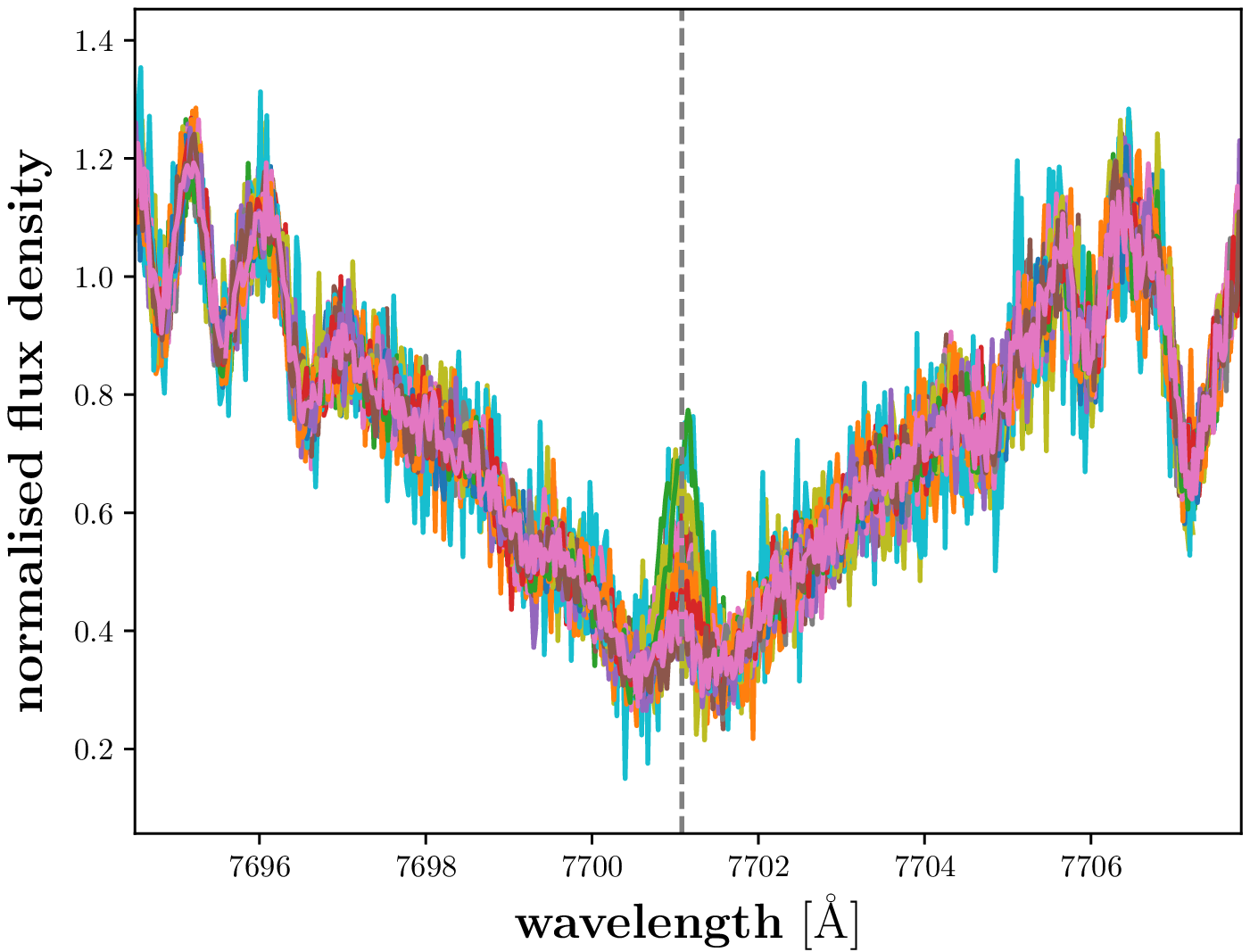}\\
\caption{\label{timeseries}  DX~Cnc time-series (\emph{top}) and correlation plot (\emph{middle})
        for \koptr\ (red circles)
        and \koptb\ (blue circles), and all spectra around the central wavelength of the \koptr\ line, marked with a dashed line (\emph{bottom}).}
\end{center}
\end{figure}

A pertinent example of correlation between the \koptr\, and H$\alpha$ line indices is observed in the
M6.5\,V star DX~Cnc (Karmn: J08298+267), shown in Fig.~\ref{timeseries}. As may be expected,
this late-type M star displays H$\alpha$ emission along with an
emission core in the \koptr\, lines. Additionally, there is variation in \kopt\, (Fig.~\ref{timeseries}). While the values of pEW(\koptr) correlate
very well with pEW(H$\alpha$), the correlation between pEW(\koptb) and this latter is weaker. The values
of the Pearson's correlation coefficient are 0.97 and 0.57 with $p$-values of $10^{-13}$ and 0.004,
respectively.
In contrast, neither the red nor blue component of the \kir\, lines shows
a significant correlation with H$\alpha$, yielding $p$-values larger than 0.4.
We show the time-series, the correlation plot, and the spectra of the
\kir\ lines for DX~Cnc in Fig. \ref{timeseriesir}.

To study correlations in the whole sample,
we first screen the time-series based on the correlation coefficient
between pEW(\koptr) and pEW(H$\alpha$).
In particular,
we consider index time-series with a Pearson's correlation coefficient of
$ |r| > 0.4$ and
$p < 0.005$ to be good candidates for correlated evolution.
A total of 59 stars  match this criterion.

In a next step, we review the time-series matching the criterion by eye.
This visual inspection
revealed that out of the 59 candidates,
7 can be attributed to outliers in the time-series. 
The MADs for the 52 stars with visually confirmed correlation are plotted
in Fig.~\ref{tellcor}. 

For the \kir\ lines, we find only five stars where both lines correlate
with pEW(H$\alpha$).
Again, visual inspection shows that one
is attributable to outliers in its time-series.
One of the four remaining stars (AD~Leo) shows
positive correlation and three stars (Ross~1020, GJ~1254, and Ross~271)
show negative correlation. All four stars show the same behaviour in the \koptr\ line.
Overall, we find about ten times fewer examples of correlation or anticorrelation for the \kir\, lines
compared to the \koptr\, line, indicating their relative insensitivity to chromospheric activity.

For both, the \kopt\
and \kir\ lines, examples of positive correlation and
anti-correlation are found. For the \koptr\, and the \kir\, lines, we
find that anti-correlation is typical for weakly variable stars, defined here
 as those with MAD(H$\alpha$) < 0.07\,\AA. This corresponds to the apparent gap in
Fig.~\ref{tellcor}, which also roughly separates active stars with H$\alpha$
in emission from stars with H$\alpha$ in absorption.
For stars with H$\alpha$ in emission we find only positive correlation,
while for stars with median(pEW(H$\alpha$)) > --0.6\,\AA\
we find three stars (Teegarden's star, GJ~1093, and Ross~619) that nevertheless exhibit a positive
correlation. This may be related to the late spectral type of these
stars (M4.5--7.0\,V), because all correlating stars with these spectral types
show a
positive correlation.
\begin{figure*}[h!t]
\begin{center}
\includegraphics[width=0.5\textwidth, clip]{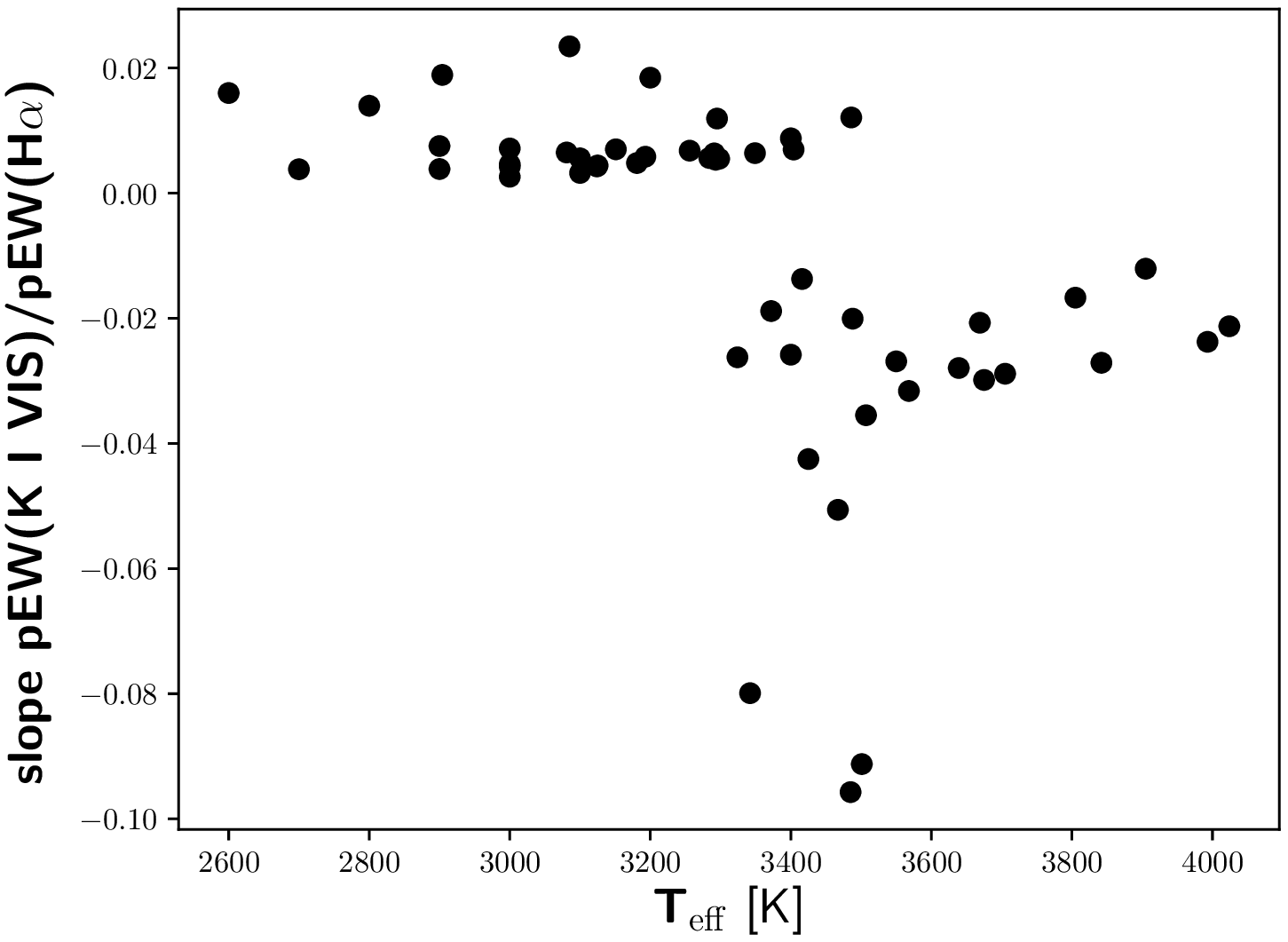}
\includegraphics[width=0.5\textwidth, clip]{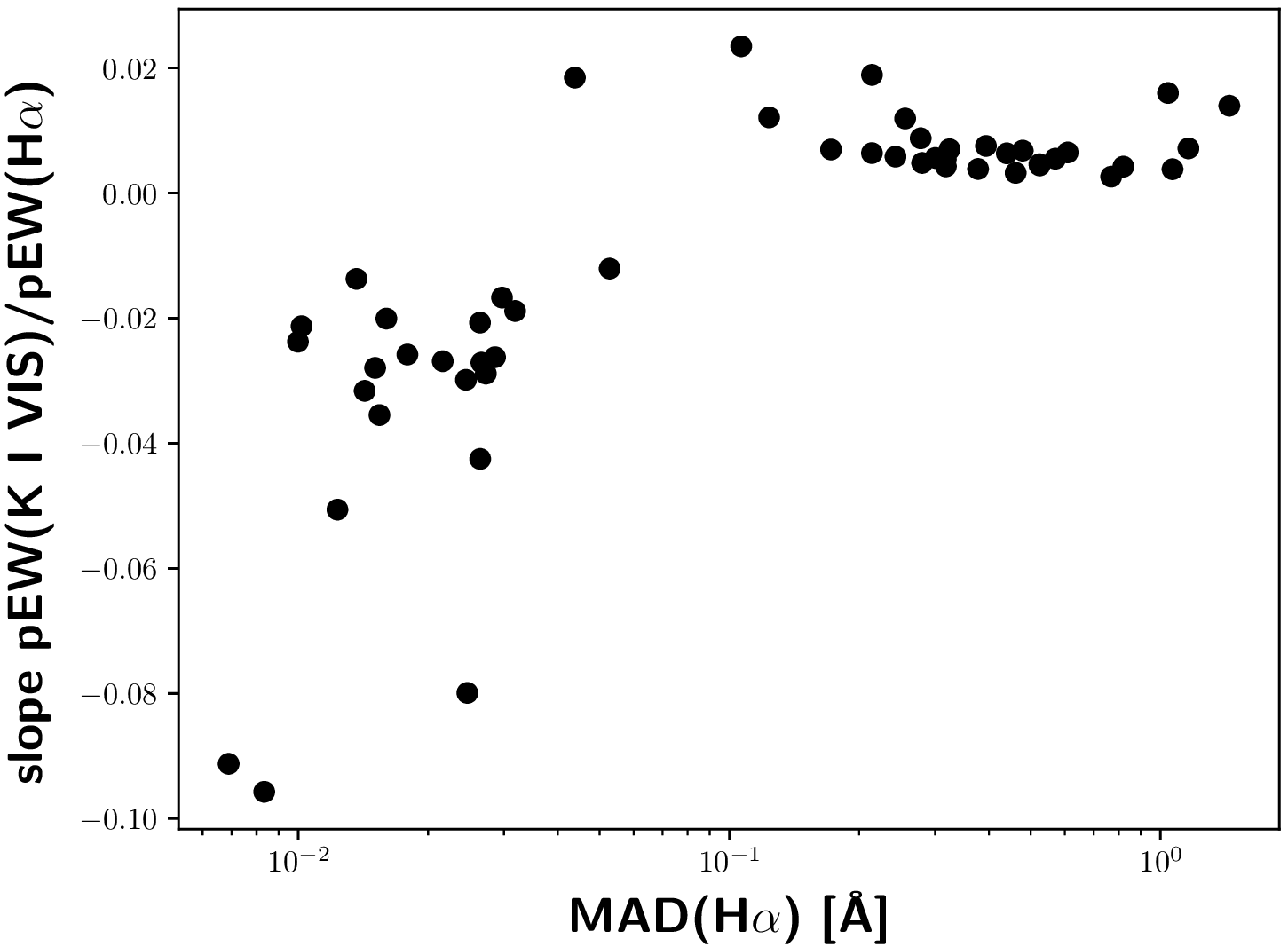}\\
\caption{\label{slopeteff}  Slope of the linear regression of pEW(H$\alpha$) and
  pEW(\koptr) as function of effective temperature (\emph{left}) and of MAD(H$\alpha$)
  (\emph{right}).
}
\end{center}
\end{figure*}

The observed behaviour agrees well with the
results of \citet{Kafka2006} and \citet{Robertson2016}, who found that early-type
inactive M stars show an anti-correlation between pEW(\kopt) and pEW(H$\alpha$), while active late-type M dwarfs
show a positive correlation. For the latter stars, H$\alpha$ is normally
in emission and higher activity corresponds to fill-in or more emission in the 
emission cores of the \kopt\ lines. For the early-type inactive stars, H$\alpha$
is normally in absorption and a fill-in is apparently accompanied by deepening of the \kopt\ lines.

The least active among 
the active stars do not show a correlation
between these two lines, which we attribute to a lower sensitivity of the
\kopt\ lines to changes in activity compared to H$\alpha$.
Therefore, we computed a linear regression to determine the slope between
pEW(\kopt) and pEW(H$\alpha$). While for the early-type stars, we find
slopes of between $-0.01$ and $-0.10$, the majority of the later type stars have
slopes of between $0.003$ and $0.01$ with seven stars
showing significantly higher slopes of between 0.01 and 0.02.
While the fits of the slopes appear statistically meaningful, the cause for the steeper slopes
remains elusive. For the four stars showing a correlation between pEW(\kir) and pEW(H$\alpha$),
we find comparable slopes between these lines.

The relation between the slope (of the linear regression between
pEW(H$\alpha$) and pEW(\koptr)) and $T_{\rm eff}$ can be seen
in Fig. \ref{slopeteff}, where the
transition from negative to positive slope takes place at $T_{\rm eff}\sim 3400$\,K
corresponding to spectral types of M3.0\,V -- M4.0\,V.
This transition from negative to positive slope may be explained by the different
origin of the H$\alpha$ and the \kopt\ lines. While H$\alpha$  originates in
plages in the upper chromosphere or even the lower transition region, the \kopt\
cores should originate in the lower chromosphere, as was found for the \ion{Na}{i}~D
lines by \citet{Andretta1997}. If the \kopt\ core originates as an absorption line in
spots, and
higher activity levels lead to an increase in the area of spots, this would
lead to a deepening line. In contrast, higher activity levels lead to more
H$\alpha$ emission in the upper chromosphere causing a fill-in of an
H$\alpha$ absorption line and therefore explaining the anti-correlation with
the \kopt\ line.
For the late-type stars, H$\alpha$ becomes an emission line, which is collisionally
controlled. The same should be true for the \kopt\ line, because for the late stars, which exhibit \kopt\ in emission, the
densities in the lower chromosphere should be quite high.
As the photospheric background is also
low, this emission shows up as emission cores correlated with H$\alpha$ emission.

The right panel of Fig. \ref{slopeteff} also shows that the slope of the
linear regression depends on the variability of the star as measured
by MAD(H$\alpha$). For anti-correlation, the slope is flattening for higher
variability, while for correlation the slope is much less dependent on the
MAD(H$\alpha$) and seems to be saturating at around 0.02.

The above findings indicate that  \kopt\ lines, and in particular the \kir\ lines,\ are less sensitive to activity than H$\alpha$. Surprisingly, the later-type, more variable stars with H$\alpha$ in emission show an even
lower sensitivity of \kopt\ compared to H$\alpha$ than stars with H$\alpha$ in absorption.
This may be caused by a higher sensitivity of H$\alpha$ emission lines to changes in the chromosphere compared to H$\alpha$ absorption lines.

For the 16 stars of spectral type M6.0\,V and later, our sample unfortunately only
comprises three stars with sufficient S/N in H$\alpha$   in the individual spectra to
compute pEW(H$\alpha$). Therefore, we cannot compute the correlation between
pEW(H$\alpha$) and pEW(\koptr) for most of these  stars. 
As they all show emission cores in the \koptr\ lines, we nevertheless assume
that for all of these stars a positive correlation exists. If this is true, the favourable
redder location of the  \kopt\ lines can make the pEW(\kopt) a valuable substitute
for pEW(H$\alpha$).

Also, we caution that we do not identify  all stars with a correlation between pEW(H$\alpha$) and pEW(\koptr),
because some stars fail our correlation criterion because of outliers in their time-series. These outliers may 
even be caused by flaring activity, where the \koptr\ line may exhibit a
different reaction to the flare than H$\alpha,$ therefore diluting the correlation.
An example is the M5.5\,V star GJ~1002, where two flares weaken
the correlation seen in the \koptr\ line. Another reason for not finding
existing correlations in the inactive stars may also be the constant activity
level of these stars during observations, when even the H$\alpha$ line shows
no variation. This latter case applies to many of 
our sample stars. In particular, we
deem all stars with MAD(pEW(H$\alpha$)) $<$ 0.01\,\AA\ to be too constant to find a
correlation. For the two stars in this regime, where we
nevertheless find an anti-correlation,
the slope is extraordinarily low  
and they both have more than 80 observed spectra.
There are 67 (20\%) stars in this low-variation regime.

\subsubsection{Variability during flares}\label{emisscores}

Due to the observing scheme of CARMENES, which is optimized for planet
detection, consecutive spectra exist so rarely, that the snapshot character
of most of the spectra prevents
us from identifying the characteristic exponential decay seen in flare light curves.
  This makes flare detection rather challenging.
  Nevertheless, large flares can be detected as `outliers' in H$\alpha$ or by asymmetries in the
  line shape as
demonstrated for example by \citet{asym, Hevar}. During these flares,
the \kopt\, lines shows fill-in. Only for the M5.0\,V star
1RXS~J114728.8+664405 (Karmn: J11474+667) is a clear
emission core identified. For the most part, it is the slightly cooler stars with spectral type M5.5\,V that 
start to develop emission cores in the \kopt\, lines also in quiescence. 
As shown in \citet{cnleoflare} using VLT/UVES observations, CN~Leo also exhibits an emission core in the \kopt\ lines 
during a mega flare, but not during quiescence also covered by these data. In our CARMENES observations, CN~Leo shows an emission
core in the average spectrum, but not in every individual spectrum, which indicates that
the \kopt\, emission is variable. 

\begin{figure*}[h!]
\begin{center}
\includegraphics[width=0.5\textwidth, clip]{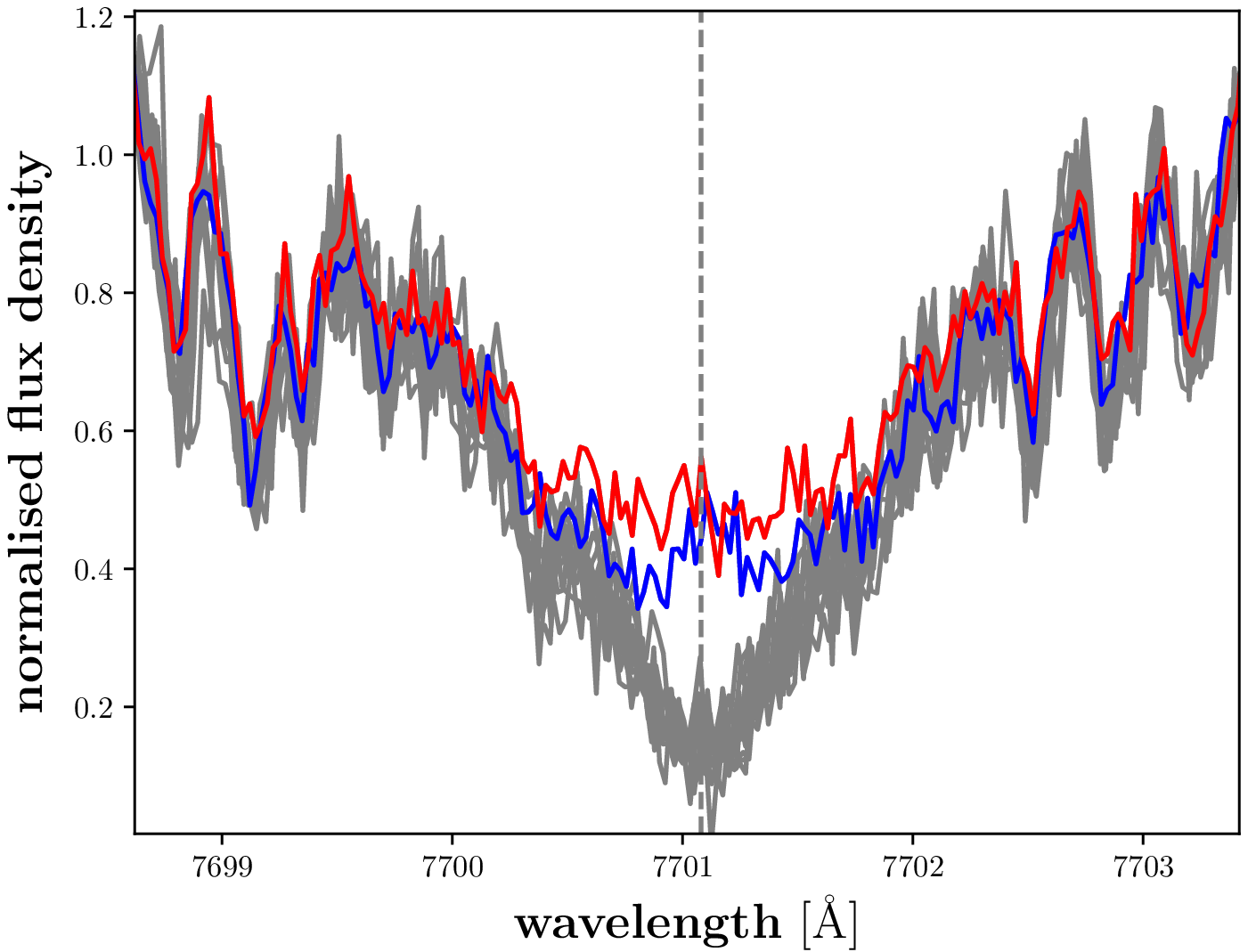}
\includegraphics[width=0.5\textwidth, clip]{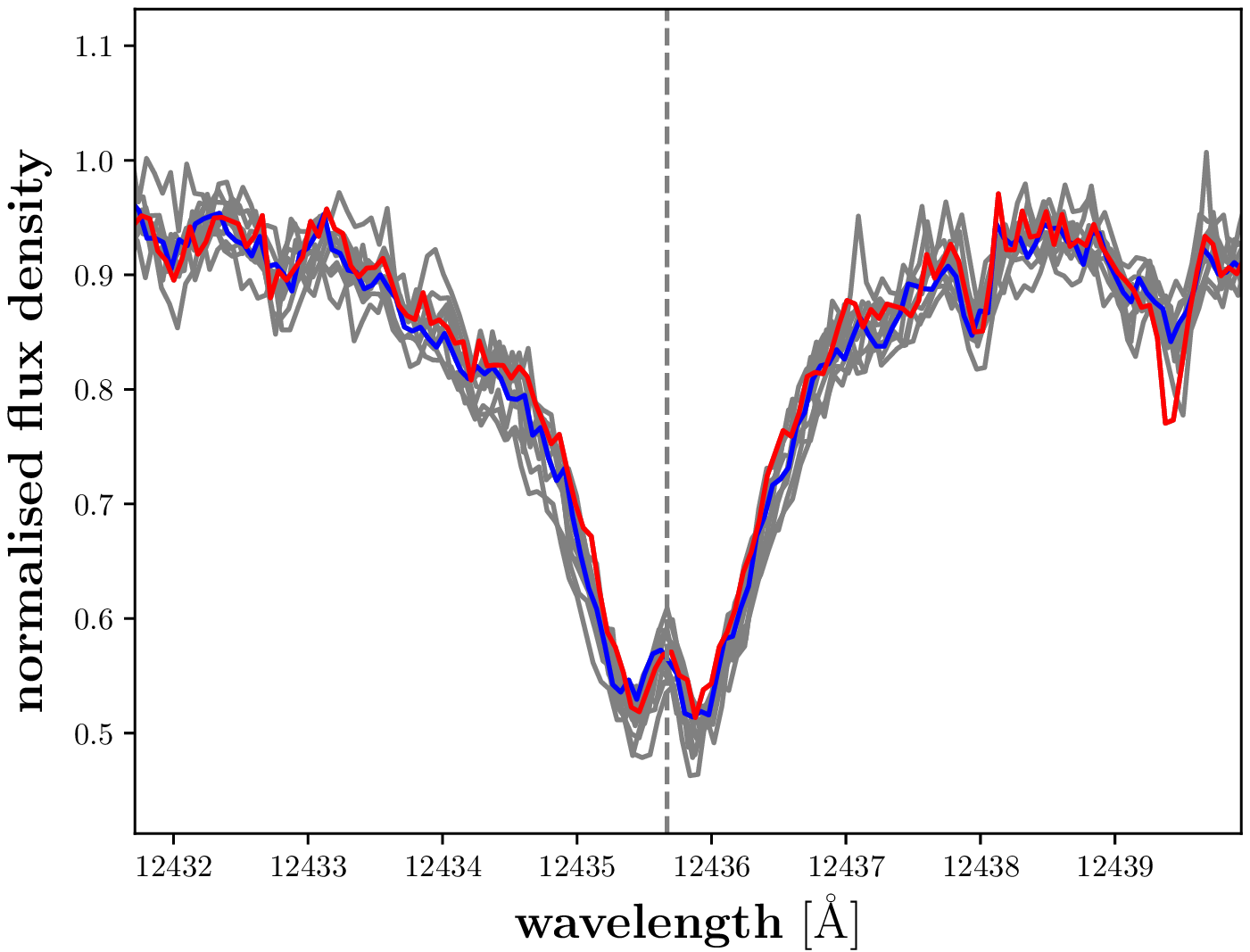}\\
\includegraphics[width=0.5\textwidth, clip]{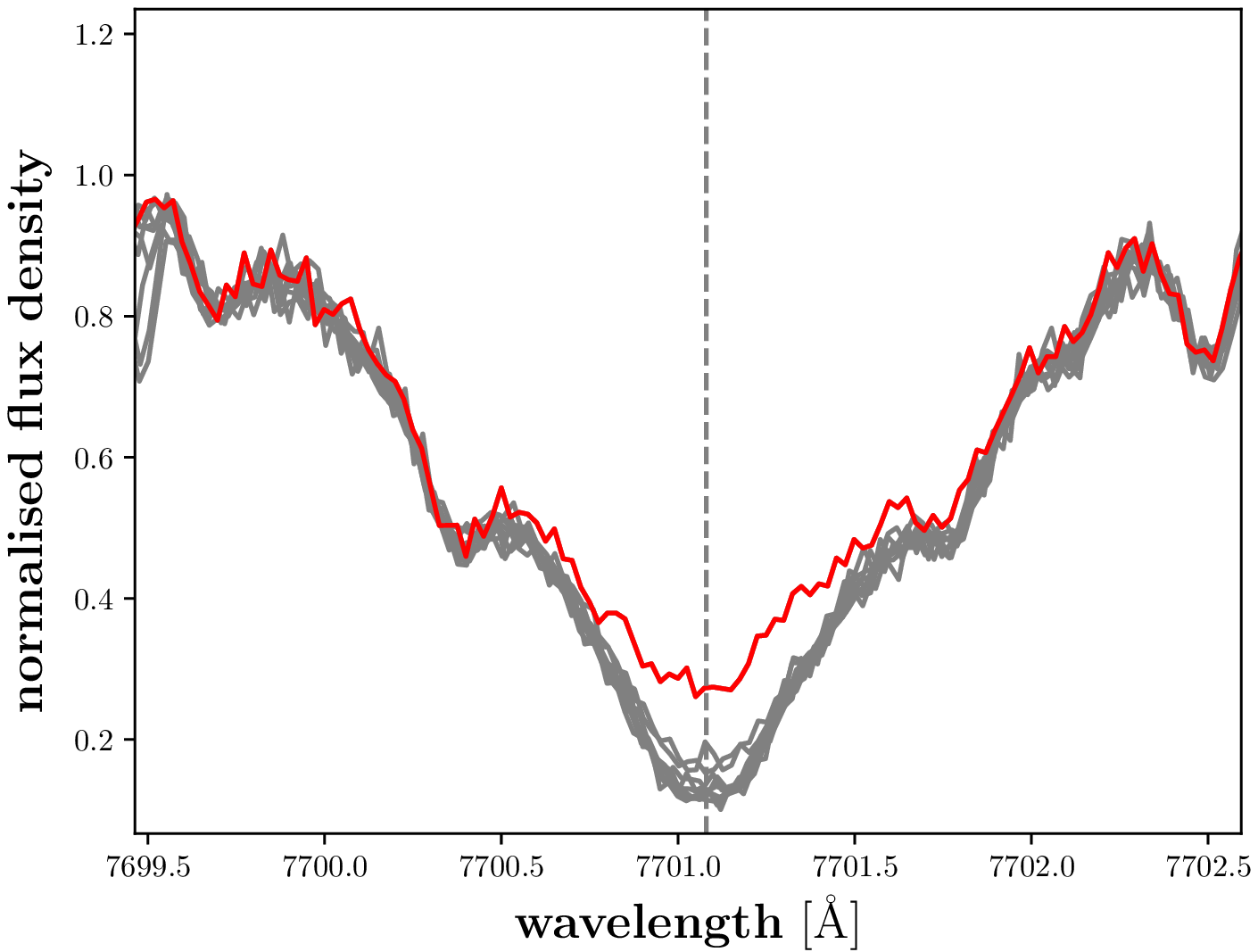}
\includegraphics[width=0.5\textwidth, clip]{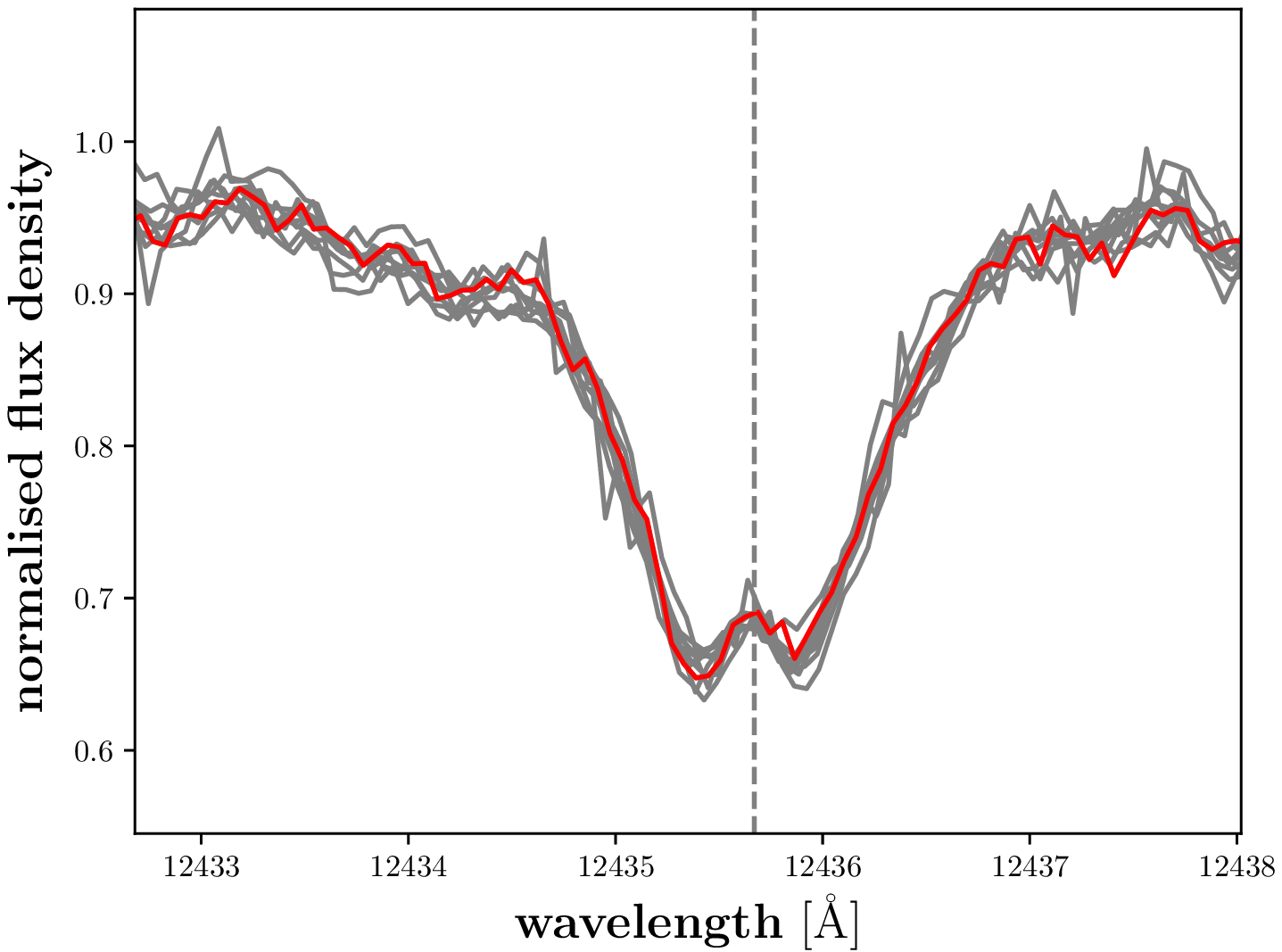}\\
\caption{\label{flares}  \koptr\ (\emph{left}) and \kirb\ (\emph{right}) spectra of
  1RXS~J114728.8+664405 (\emph{top}) and 
EV~Lac  (\emph{bottom}) during quiescence (grey lines) and flares (red and blue lines).
In contrast to the \koptr\ line, the \kirb\ line does not appear to react to flares.
The double dip structure of the \kirb\ line is explained in Sect. \ref{Zeeman}.
}
\end{center}
\end{figure*}

We show two instructive examples of flare spectra in Fig.~\ref{flares}. In the bottom
panel, we display  some of our spectra of
EV~Lac (Karmn: J22468+443), including the spectrum taken at JD 2457633.46711\,d, which corresponds
to the largest flare of this star in the time-series. 
The behaviour of other chromospheric indicator lines during this
flare has been described in \citet{asym, Hevar}, as well as the appearance 
of large
red asymmetries in H$\alpha$ and other lines.
Here, the \koptr\ line spectrum  also shows this red asymmetry, which is  most probably
caused by mass motions. 

Another example of a strong flare affecting the \koptb\ line is shown in the top
panel of Fig. \ref{flares} for spectra of 1RXS~J114728.8+664405.
One flare leads to a strong fill-in, while
the second flare leads to an emission core. How this different reaction of the line shape
is caused remains unknown: H$\alpha$ exhibits strong red
asymmetries for both flares, with the main difference being the amplitude of the H$\alpha$ line.
Comparison spectra of H$\alpha$ and the \ion{He}{i} IR line can be found again
in \citet{Hevar} for the flare marked in blue in Fig. \ref{flares}. The other
flare has been observed more recently and is not included in the study by \citet{Hevar}.

In contrast to the \kopt\, lines, no clear variability is found in the \kir\, lines,
not even during the largest flares in EV~Lac or 1RXS~J114728.8+664405
as can be seen for EV~Lac in Fig.~\ref{flares}. While the redder of the absorption dips of the line shows some deformation during the flare, the bluer dip does
not. Though we cannot rule out some reaction to the flare, we think this is most probably noise, because the line centre does not show any variation.

This non-reaction to flares strongly suggests that the correlation or anti-correlation found between the pEW(\kir)
and pEW(H$\alpha$) for the four stars mentioned above is caused by some alternative mechanism to 
that  causing the correlation between pEW(\koptr) and pEW(H$\alpha$). Possible alternative mechanisms
are: ($i$) The \kir\ lines may originate not from the chromosphere, but  mainly
from photospheric spots, whose relative area should also be roughly correlated with
H$\alpha$ activity but would not change during flares. ($ii$) As the \kir\ lines are magnetically sensitive as detailed
in the following section, changes in the magnetic field may change the shape of these lines
and therefore lead to variations in pEW(\kir). If pEW(H$\alpha$) correlates well with the magnetic
field, a correlation with pEW(\kir) should also appear.

\subsubsection{Zeeman effect}\label{Zeeman}

For about 20 active stars in our sample, the \kir\, lines show a double minimum structure imitating an emission core.
While this profile resembles a chromospheric emission line core,
we do not observe any flare-induced variability.
For example, the large flare on EV~Lac (see Sect.~\ref{emisscores})
leads to considerable fill-in in the \kopt\ lines, as seen
in Fig.~\ref{flares}, but does not clearly affect the profiles of the \kir\ lines.
Also, chromospheric computational models,
such as those constructed by \citet{Hintz2019} for M2--3 dwarfs using PHOENIX,
do not show any emission cores or noteworthy fill-in for the \kir\ line profiles,
even for extreme flaring models.





\begin{figure}[t!]
\begin{center}
\includegraphics[width=0.5\textwidth, clip]{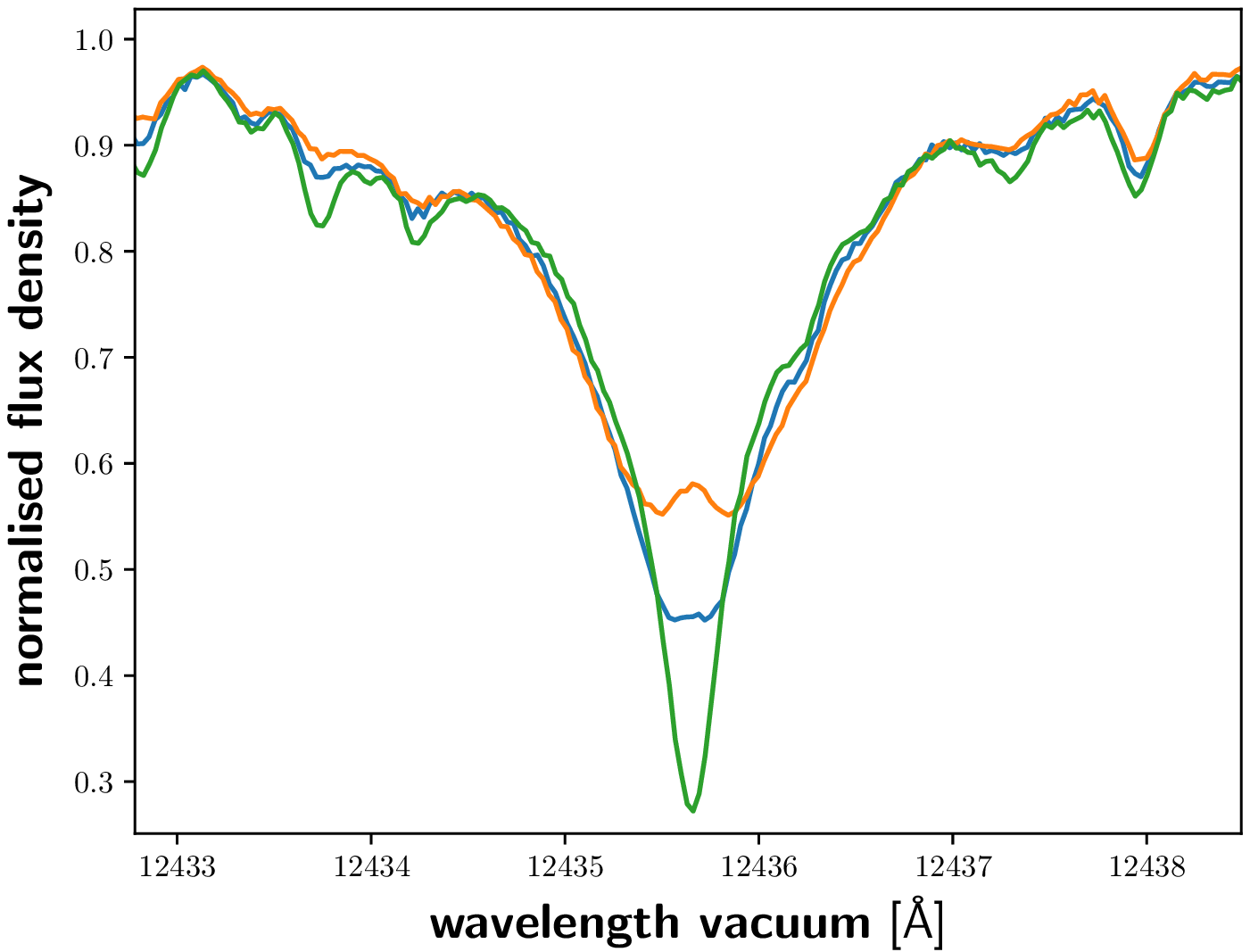}%
\caption{\label{zeemanemiss}  Spectra of the M5.0\,V stars G~177-025 (blue), LP~731-076
  (orange), and GJ~1028 (green), depicting the different line shapes found (flat bottom,
  double dip structure, and unperturbed).
}
\end{center}
\end{figure}

These considerations lead us to the hypothesis that the line profile is caused by 
magnetic Zeeman splitting. This would also explain that
for nearly all stars with a double dip structure, the line profile of the
\kir\ lines is slightly broadened compared
to inactive stars of the same spectral type, which 
show a `normal' single dip line-shape.
Also, in comparison to the chromospheric models by \citet{Hintz2019}, who treats
\ion{K}{i} in non-LTE, the
observed line profiles are broader than the profiles of the model spectra.
Additionally,  \citet{Shulyak2019} also
  used CARMENES spectra to determine magnetic fields in 29 stars using
  direct magnetic spectrum synthesis. Five of the stars, where we identify
  the peculiar double dip structure are also part of this study. We find
  for these five stars that the distance between the two dips grows with
  measured magnetic field.

Finally, we found the line profile of the  \kir\ lines
to have a `flat bottom' (a trough-like shape) in a number of stars. 
Examples are the stars LTT~11262, 
2M~J07000682-1901235, and G~177-025. Typical examples of a flat bottom, a double dip
structure, and an unperturbed line can be found in Fig. \ref{zeemanemiss}.
Measurements
  by \citet{Shulyak2019}  for two stars with a flat bottom line shape show that their magnetic fields are of approximately the same magnitude as the stars with the double dip structure, but the same two stars
  have larger \vsini\ values in comparison. Therefore,
the flat bottom may indicate unresolved Zeeman
splitting, which represents the typical case for M dwarfs, while the double dip structure
is due to resolved Zeeman splitting. 
Resolved Zeeman splitting
has been used more frequently to measure field strengths in chemically peculiar early-type stars
\citep[e.g.][]{Chojnowski2019}.

\begin{figure}[h!]
\begin{center}
\includegraphics[width=0.5\textwidth, clip]{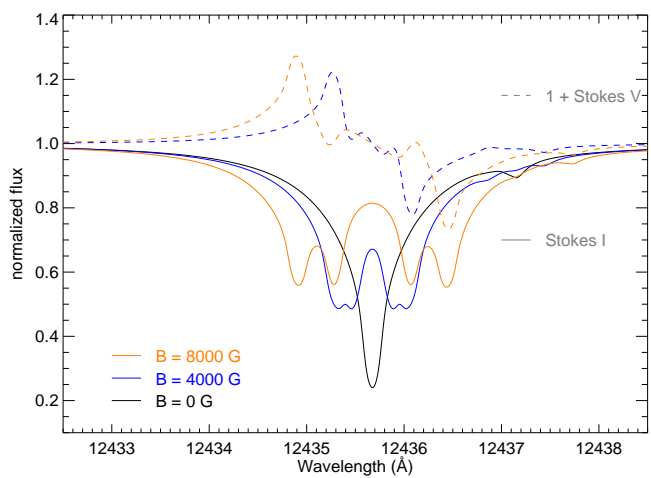}\\
\caption{\label{Stokes} Stokes I (solid lines) and Stokes V (dashed lines)
  for the \kirb\ line for different magnetic field strengths computed for a
        purely radial field. }
\end{center}
\end{figure}

In the case of resolved Zeeman splitting,
the (half-)width of magnetic line splitting, $\Delta \lambda$, is characterised
by the Land\'e factor, $g_i$.
In particular, the magnetic field strength $B$ in kG is related to the split by
\begin{equation}
        \Delta\lambda_i = B \cdot 4.67\cdot 10^{-13} \AA^{-1} G^{-1} \cdot \lambda^{2} \cdot g_{i}
\label{eq:B}
,\end{equation}
where $\Delta\lambda$ is the split in wavelength in \AA\,\citep{Reiners2012}.
Therefore, NIR lines show greater splitting because of the quadratic dependence
on wavelength. To compute the magnetic field $B$, the Land\'e factor for the line
transition is required, but no experimental
  Land\'e factors are known for the \kir\ line. 
Land\'e factors can be estimated with
the formula given by \citet{Reiners2012}:
\begin{equation} g_{i}=\frac{3}{2}+\frac{S_{i}(S_{i}+1)-L_{i}(L_{i}+1)}{2J_{i}(J_{i}+1)} \; . \end{equation}
Here, $S$ is the spin, $L$ is the orbital angular momentum, and $J$ is the total angular 
momentum quantum number, and $i$ denotes the level (here upper and lower).
Here, the two
energy levels of the transition each split into two components because of
spin-orbit coupling, i.e. the anomalous Zeeman effect. The Land\'e factors
of the upper and lower energy levels for the \kirb\ line are $g_{\rm up} = 2.0$ and
$g_{\rm low} = 0.67$, and for the \kirr\ line $g_{\rm up} = 2.0$ and $g_{\rm low} = 1.33$,
respectively. In this case, the Zeeman pattern of the
spectral line only consists of components that are shifted with respect to
the undisturbed wavelength, but no undisturbed component remains. The
resulting pattern for the \kirb\ line is shown in Fig.~\ref{Stokes} for different field
strengths and a radial field geometry. Because all Zeeman components are shifted, the pattern develops
an inversion in the centre at field strengths larger than $\sim 1$kG. This
inversion is reminiscent of central line emission and is more pronounced in
stronger fields.

The \kopt\ doublet also shows a similar pattern of Land\'e factors, which
would lead
in principle to a double dip structure in these lines as well. 
However, the shorter wavelength of the \kopt\ lines leads to a much narrower
Zeeman pattern, which prevents any observations. 

The exact evaluation of the magnetic field should be performed
using detailed radiative transfer calculations, which is beyond the scope of this paper.
We nevertheless attempted some forward modelling of the Zeeman effect
using the spectrum synthesis code {\tt MAGNESYN} \citep{Shulyak2017}, which solves the radiative transfer
equations for all four Stokes parameters and a given magnetic configuration.
The code is part of the {\tt LLMODELS} suite described in \citet{Shulyak2004} and
was used by 
\citet{Shulyak2019} to derive magnetic fields for CARMENES spectra.

\begin{figure}[h!]
\begin{center}
\includegraphics[width=0.5\textwidth, clip]{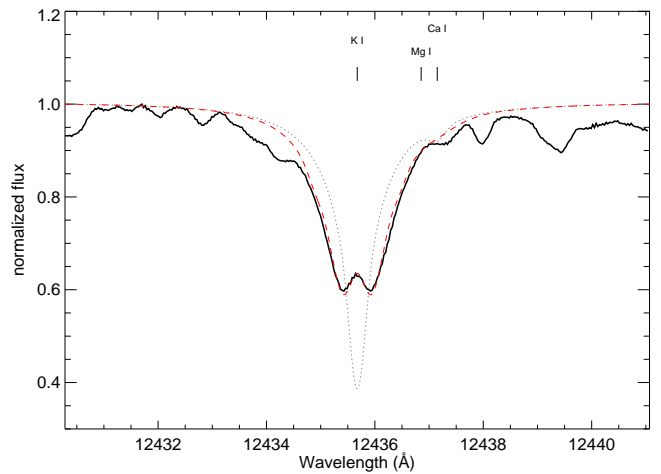}\\
\caption{\label{forwardmodel} Forward modelling of the Zeeman splitting
        in the \kirb\ line. Shown is the observed spectrum of YZ~CMi in black.
        The dotted grey line is a radiative transfer model without
        magnetic field. The red dashed line has been computed with 
        a magnetic field consistent
        with that found by \citet{Shulyak2019}.
        Atomic lines used in the models are marked at the top. }
\end{center}
\end{figure}

We show an example of forward modelling for the \kirb\ line in Fig. \ref{forwardmodel}
for the star YZ~CMi (Karmn: J07446+035). 
For this star, the magnetic field was determined by \citet{Shulyak2019}
to be $<B>$ = 4.6 $\pm$ 0.3\,kG with a weak zero-field component.
Using a model consistent with the findings by \citet{Shulyak2019} leads to
the well-fitting model shown in Fig. \ref{forwardmodel}. The same
model leads to no double dip feature in the \koptb\ line as expected
from our considerations above.

\section{Conclusions}\label{conclusion}

We present a study of the \kopt\ and \kir\ doublet lines in a sample of 324 M dwarfs, using
a total of more than 15\,000 spectra observed
within the CARMENES exoplanet survey. Both doublets show prominent photospheric lines
across the M dwarf sequence, the profiles of
which strongly depend on effective temperature and also on gravity.

We determine pEWs for the line cores of the \kopt\ and \kir\ lines in
band passes optimised for the study of chromospheric activity, which also allow us
to deduce relations between pEW, $T_{\mathrm{eff}}$, and $\log\,g$, for slow rotators with \vsini $<$ 15\,km\,s$^{-1}$.
As the pEW measurements for the blue line component of the \kopt\ doublet
can be affected by telluric correction artefacts,
we only consider the \koptr\ line for variability studies.


We find that stars later than spectral type M5.0\,V exhibit an emission core in
the \kopt\ lines. Throughout the M dwarf sequence, we find 52 (16\%) stars that show a close correlation or
anti-correlation
between pEW(\kopt) and pEW(H$\alpha$). While stars with stronger variability ---which normally also show
H$\alpha$ in emission and are of spectral type later than about M3.0\,V--- tend to show a positive correlation between
pEW(\kopt) and pEW(H$\alpha$), less-variable, inactive early-type stars show an anti-correlation.
This is in agreement with the findings of \citet{Kafka2006}, who found an anti-correlation between
pEW(\kopt) and pEW(H$\alpha$) for early M dwarfs, and \citet{Robertson2016}, who found a correlation
for two late-type M dwarfs.
For the \kir\ lines, we also find four (1\%) cases of correlation or anti-correlation with H$\alpha$.
The potassium lines are significantly less sensitive to chromospheric variability
than H$\alpha$, and for the more active stars with H$\alpha$ in emission the difference
seems to be even more severe, which we ascribe
to an increased sensitivity of the pEW(H$\alpha$)  to variability for these stars.
This low sensitivity of the \kopt\ lines to activity-related variability in comparison to H$\alpha$
---which is particularly pronounced for
the \kir\ lines--- favours their usage in transmission
spectroscopy of exoplanetary atmospheres. Also, the magnetic splitting
pattern detected for the \kir\ lines did not show significant changes in our 
data, especially not during flares, where changes in the magnetic field due to
reconnection may be expected. Whether or not long-term variation of the magnetic field ---especially with slow rotation or cycles---  leads to 
changes in the line profiles remains to be investigated, but changes on these timescales should not affect transmission spectroscopy.  

During strong flares, the \kopt\ lines show a pronounced fill-in or development of an emission
core for spectral types around M5.0\,V. However, the \kir\ line profiles do not show a significant change 
during these transient events. Nevertheless, the \kir\ lines are sensitive to the stellar 
magnetic field. The line shape often exhibits a double dip structure for 
active stars, which
we attribute to resolved anomalous Zeeman splitting caused by a symmetric split by the $\pi$-
and $\sigma$-components.
We therefore find that the \kir\ lines are an important new diagnostic of magnetic fields in M dwarf stars.


\begin{acknowledgements}
  B.~F. acknowledges funding by the DFG under Schm \mbox{1032/69-1}.
  CARMENES is an instrument for the Centro Astron\'omico Hispano-Alem\'an de
  Calar Alto (CAHA, Almer\'{\i}a, Spain). 
D.~S. acknowledges the financial support from the State Agency for Research of
the Spanish MCIU through the "Center of Excellence Severo Ochoa" award to
the Instituto de Astrof\'{\i}sica de Andaluc\'{\i}a (SEV-2017-0709)
  We acknowledge financial support from the Agencia Estatal de Investigaci\'on
of the Ministerio de Ciencia, Innovaci\'on y Universidades and the ERDF
through projects
 PID2019-109522GB-C5[1:4]/AEI/10.13039/501100011033     
 PGC2018-098153-B-C33                                   
and the Centre of Excellence ``Severo Ochoa'' and ``Mar\'ia de Maeztu''
awards to the Instituto de Astrof\'isica de Canarias (CEX2019-000920-S),
Instituto de Astrof\'isica de Andaluc\'ia (SEV-2017-0709), and Centro de
Astrobiolog\'ia (MDM-2017-0737), and the Generalitat de Catalunya/CERCA
programme.
  CARMENES is funded by the German Max-Planck-Gesellschaft (MPG), 
  the Spanish Consejo Superior de Investigaciones Cient\'{\i}ficas (CSIC),
  the European Union through FEDER/ERF FICTS-2011-02 funds, 
  and the members of the CARMENES Consortium 
  (Max-Planck-Institut f\"ur Astronomie,
  Instituto de Astrof\'{\i}sica de Andaluc\'{\i}a,
  Landessternwarte K\"onigstuhl,
  Institut de Ci\`encies de l'Espai,
  Institut f\"ur Astrophysik G\"ottingen,
  Universidad Complutense de Madrid,
  Th\"uringer Landessternwarte Tautenburg,
  Instituto de Astrof\'{\i}sica de Canarias,
  Hamburger Sternwarte,
  Centro de Astrobiolog\'{\i}a and
  Centro Astron\'omico Hispano-Alem\'an), 
  with additional contributions by the Spanish Ministry of Economy, 
  the German Science Foundation through the Major Research Instrumentation 
    Programme and DFG Research Unit FOR2544 ``Blue Planets around Red Stars'', 
  the Klaus Tschira Stiftung, 
  the states of Baden-W\"urttemberg and Niedersachsen, 
  and by the Junta de Andaluc\'{\i}a.

\end{acknowledgements}

\bibliographystyle{aa}
\bibliography{papers}

\appendix



\section{Relation of [Fe/H] to pEW(\koptr) and pEW(\kirr)}\label{app:a}

\begin{figure}[h!]
\begin{center}
\includegraphics[width=0.5\textwidth, clip]{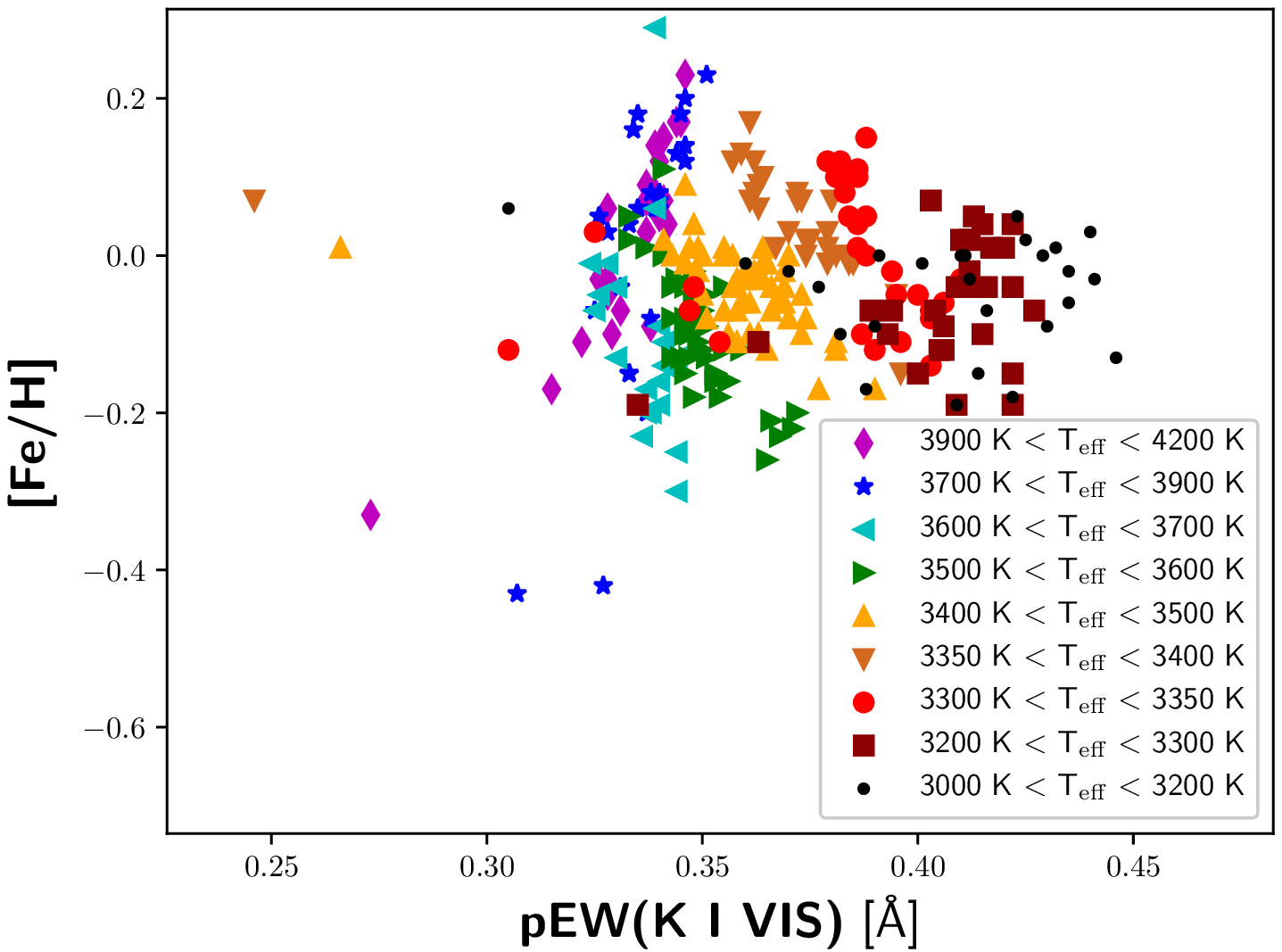}\\
\includegraphics[width=0.5\textwidth, clip]{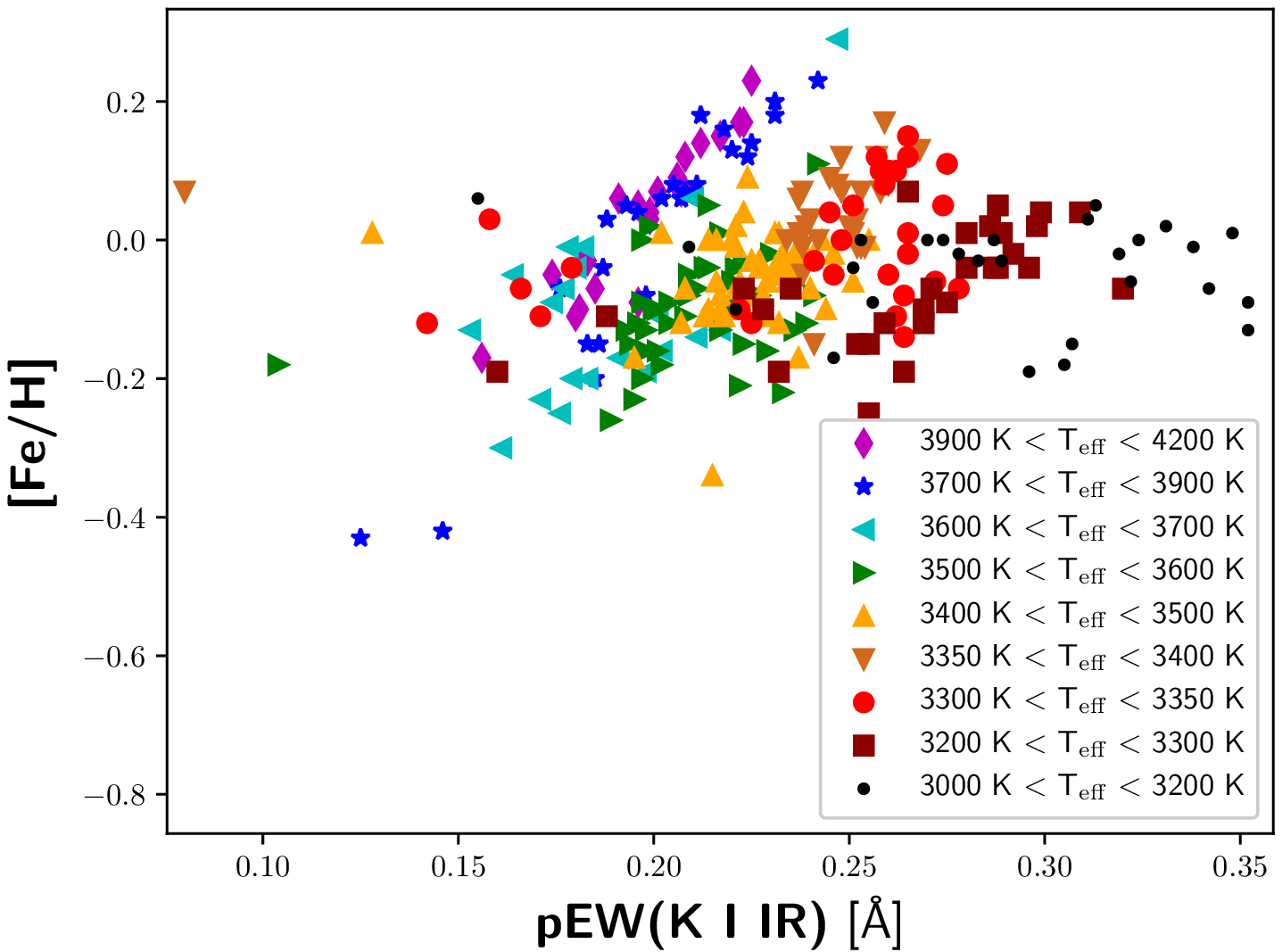}\\
\caption{\label{knirfeh}  [Fe/H] as a function of pEW(\koptr) \emph{(top)}
  and pEW(\kirr) \emph{(bottom)}. 
  }
\end{center}
\end{figure}

 Figure \ref{knirfeh} shows the distribution of [Fe/H] over pEW(\koptr) and pEW(kirr),
respectively. No relation is found.

\section{The pEW(\koptr) and pEW(\kirr) as activity indicators}\label{app:b}
\begin{figure}
\begin{center}
\includegraphics[width=0.5\textwidth, clip]{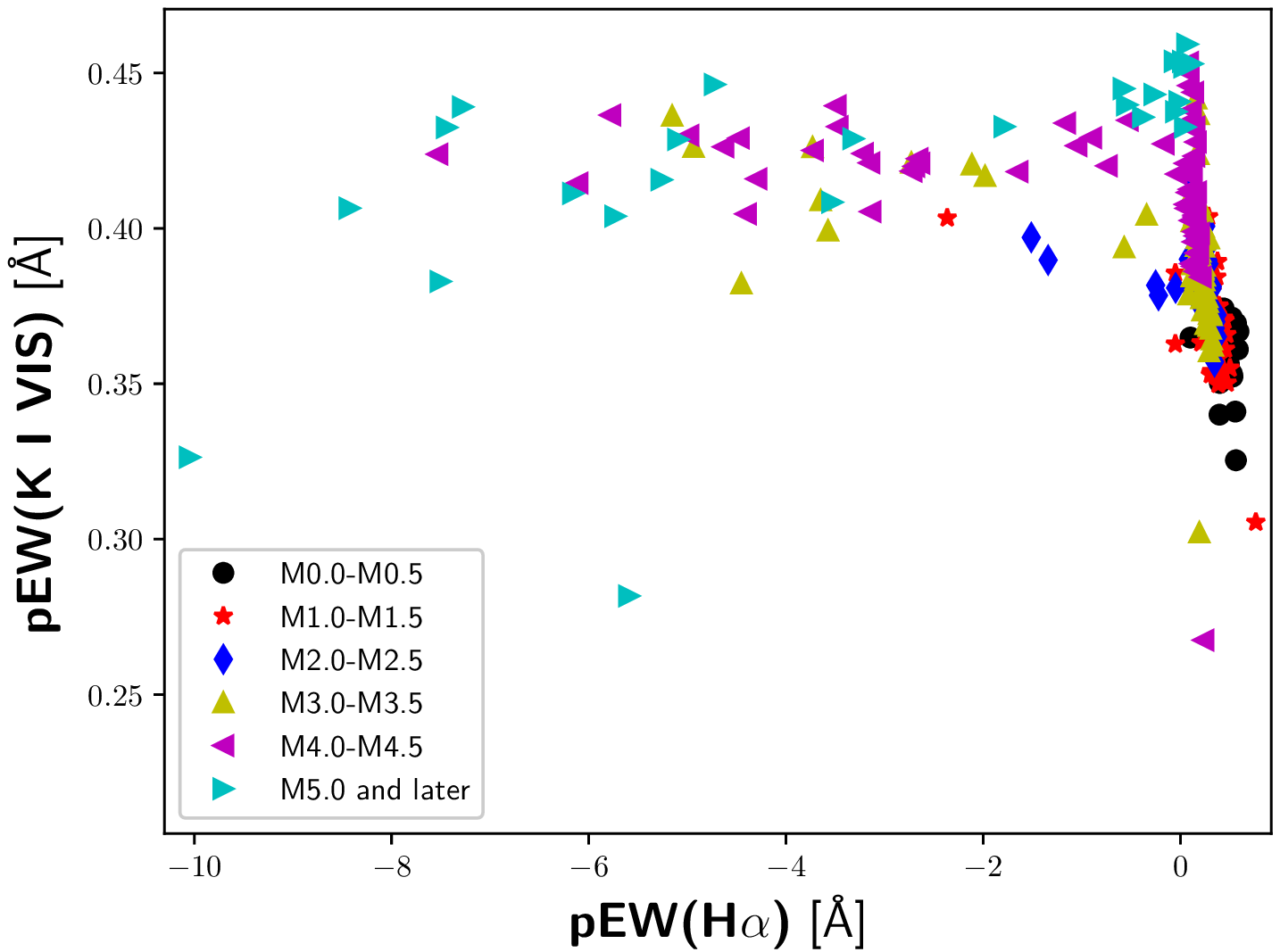}\\
{-0.4mm}\\
\includegraphics[width=0.5\textwidth, clip]{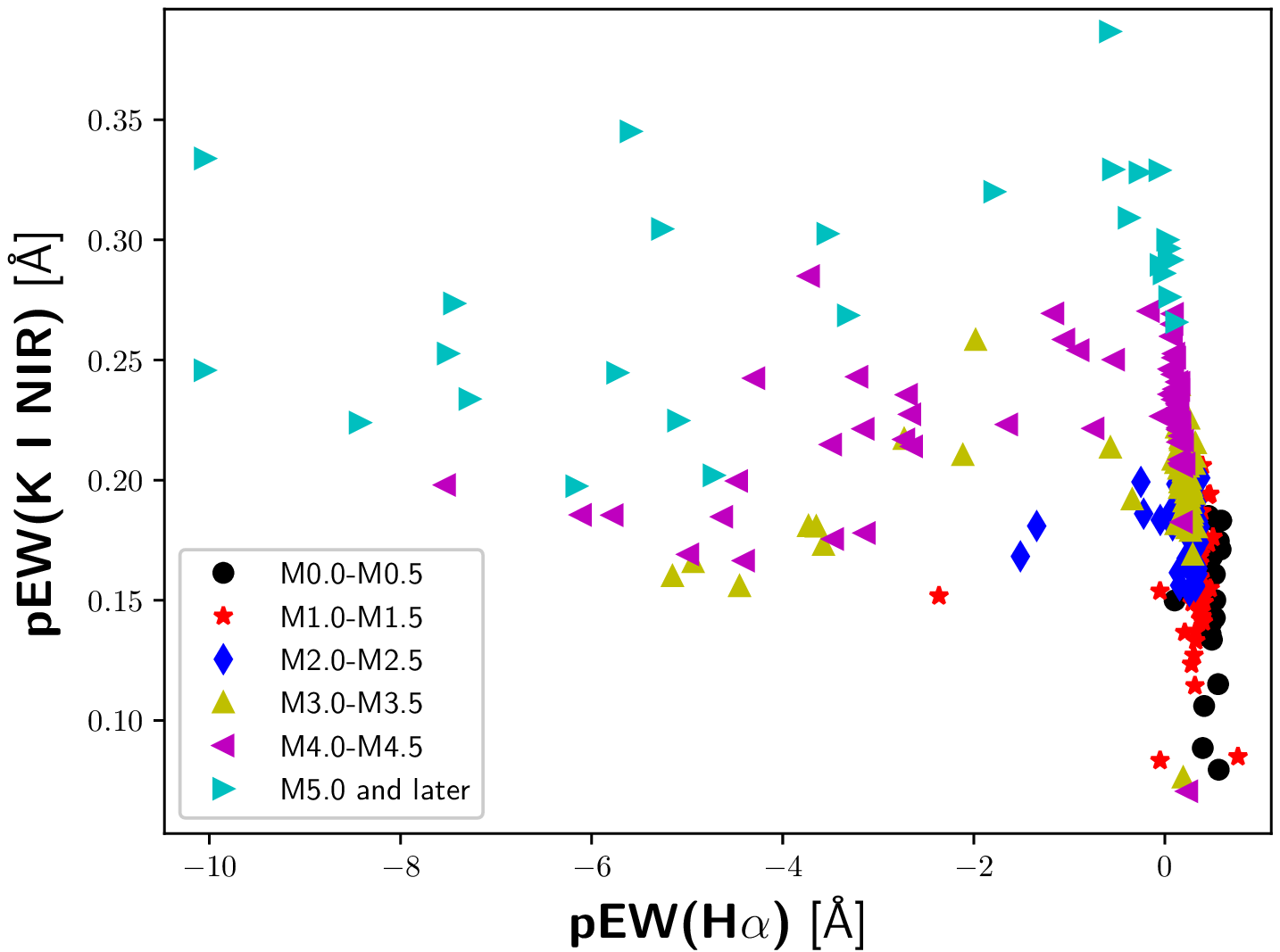}\\
\caption{\label{kopthalpha}  Dependency of pEW(\koptr) (\emph{top}) and pEW(\kir
r) (\emph{bottom})
  on pEW(H$\alpha$) for different spectra types for all stars with \vsini $<$ 15
\,km\,s$^{-1}$ .
}
\end{center}
\end{figure}

We additionally show the dependency of pEW(\koptr) on pEW(H$\alpha$)
for all slow rotators with vsini $<$ 15\,km\,s$^{-1}$  in Fig. \ref{kopthalpha}. We exclude
faster rotators because the flattening of the spectral lines by rotation would lead to
a larger effect on the pEW. For the inactive stars with pEW(H$\alpha$)
$>$ $-0.6$\,\AA,\, the M0.0\,V and M1.0\,V stars form a cloud-like structure, while later type
stars form linear segments. We therefore  argue that the dependency for these stars
is mainly caused by the temperature dependency and that pEW(\koptr) is not a good activity
tracer for such spectral types, at least not for comparison between stars. The same
applies to the \kirr\, lines for the inactive stars. Similarly, for the active stars, no clear correlation between pEW(H$\alpha$) and pEW(\koptr) is seen. Indeed the deepest
lines with the largest pEW(\koptr) are measured for the inactive stars of each spectral
subtype. While the pEW(\koptr) for the active stars seem to be approximately constant, for the
pEW(\kirr) a weak correlation is observed with pEW(H$\alpha$) indicating some sensitivity
to activity.

\begin{figure}
\begin{center}
\includegraphics[width=0.5\textwidth, clip]{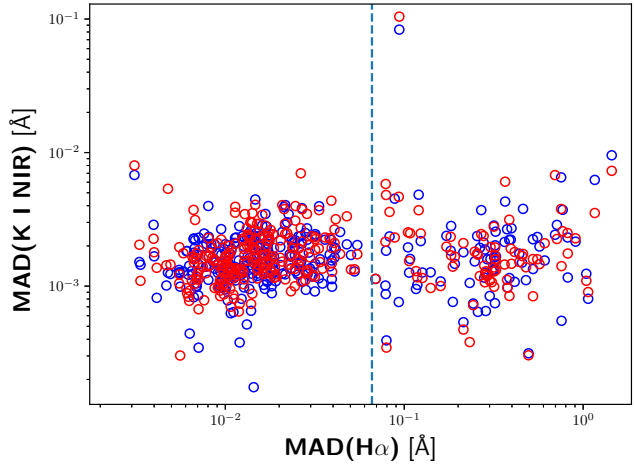}\\
\caption{\label{tellcorrir}  Same as Fig. \ref{tellcor}, but for the \kirr\ (red circles)
        and the \kirb\ (blue circles) line. The two outliers at the top are
        caused by the same star, which has only four measurements, out of which two
        happen to be compromised by low S/N, which leads to the large MAD(\kir) values.
}
\end{center}
\end{figure}

Moreover, as a comparison to Fig.~\ref{tellcor} we show in Fig.~\ref{tellcorrir} the relation between MAD(\kir) and
MAD(H$\alpha$).

\begin{figure}
\begin{center}
\includegraphics[width=0.5\textwidth, clip]{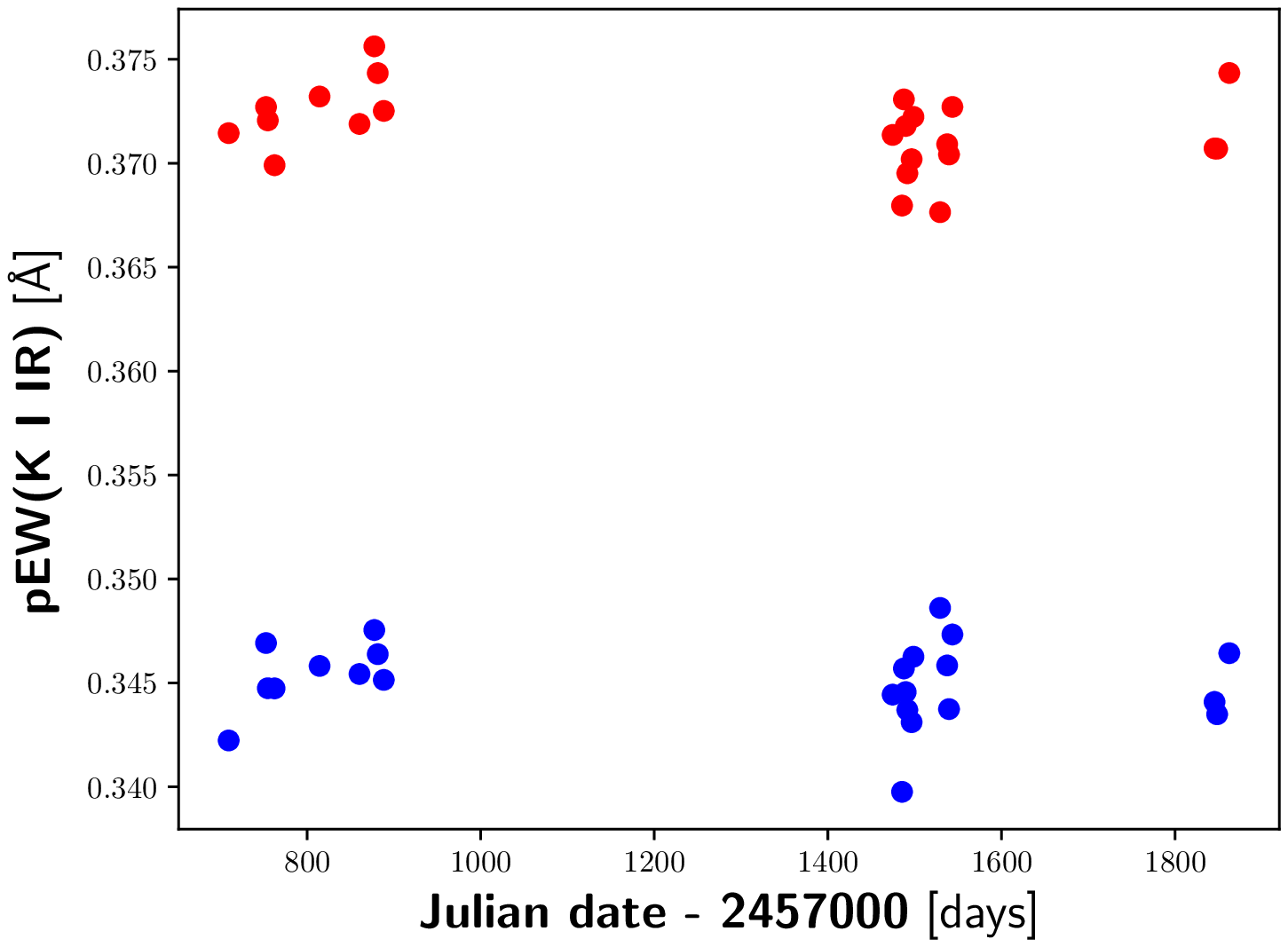}\\
\includegraphics[width=0.5\textwidth, clip]{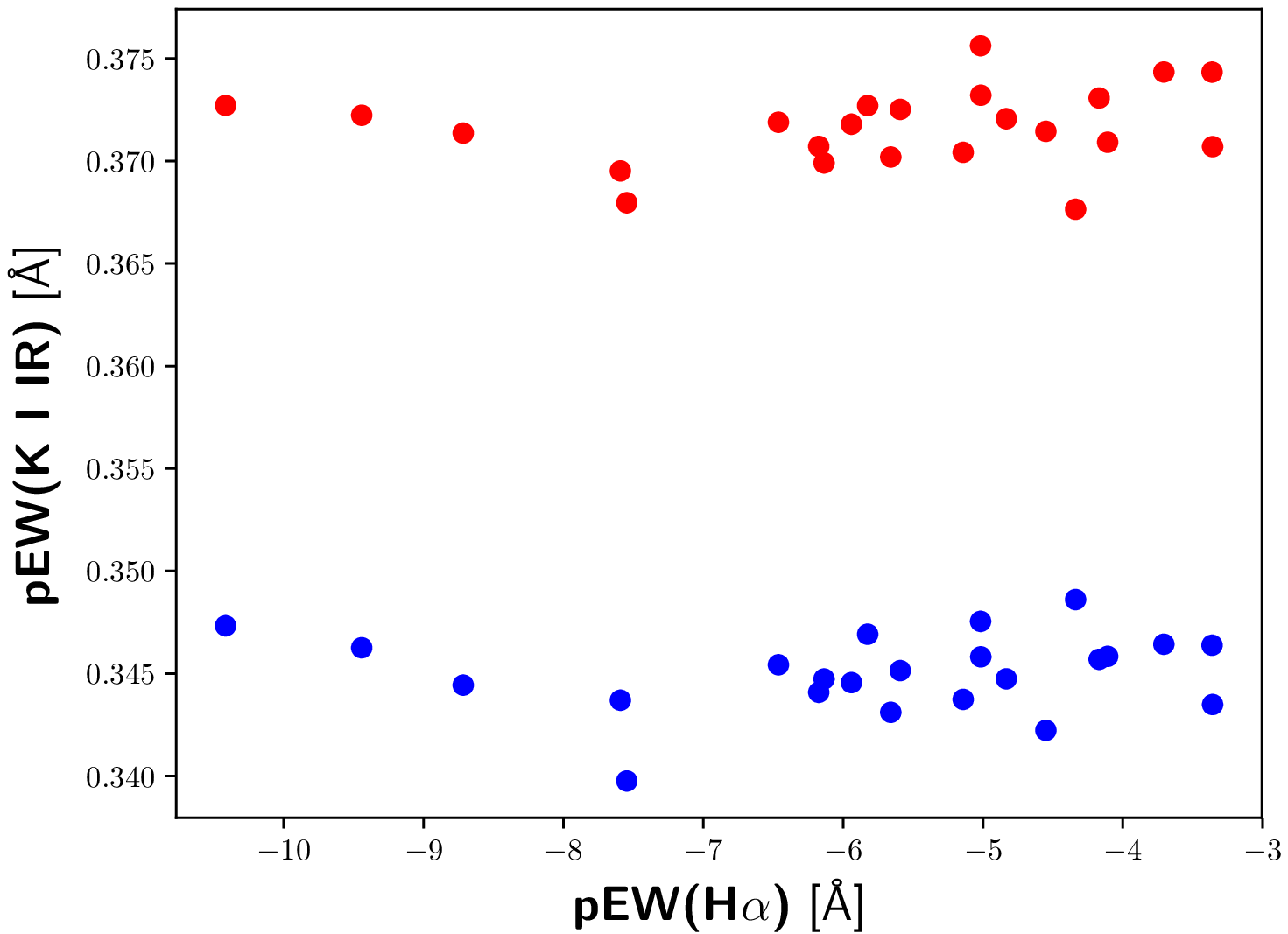}\\
\includegraphics[width=0.5\textwidth, clip]{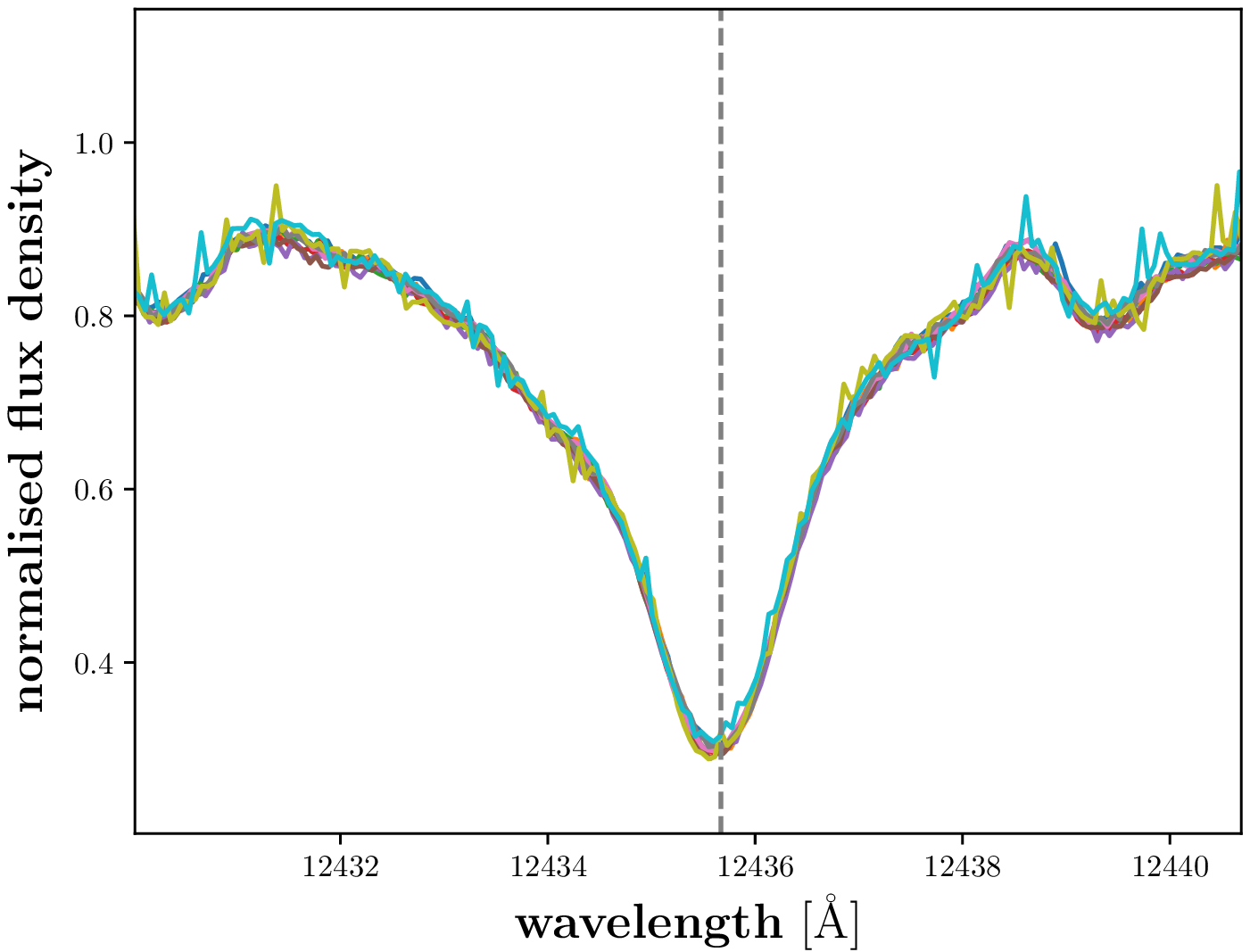}\\
\caption{\label{timeseriesir}  Same is in Fig. \ref{timeseries}, but for the
\kir\ line.}
\end{center}
\end{figure}

For comparison with Fig.~\ref{timeseries}, we show for the star DX~Cnc the time-series for the \kir\ lines, their correlation
to pEW(H$\alpha$), and the spectra of the \kirb\ line in Fig. \ref{timeseriesir}.

\end{document}